\documentclass{aa}
\usepackage{graphicx}
\usepackage{tabularx}
\usepackage{multirow}
\usepackage{subfig}
\usepackage{url}
\usepackage{amsmath}
\usepackage[dvipsnames]{xcolor}
\usepackage{hyperref}
\usepackage{cleveref}
\usepackage{soul}
\usepackage{verbatim}
\hypersetup{
    colorlinks=true,
    linkcolor=blue,
    filecolor=magenta,      
    urlcolor=cyan,
    citecolor=blue,
}
\begin{document}

\title{3D modeling from the onset of the SN to the full-fledged SNR}

\subtitle{Role of an initial ejecta anisotropy on matter mixing}

\author{A. Tutone
\inst{1}\fnmsep\inst{2}
\and
S. Orlando
\inst{3}
\and
M. Miceli
\inst{1}\fnmsep\inst{3}
\and
S. Ustamujic
\inst{3}
\and
M. Ono
\inst{4}\fnmsep\inst{5}
\and
S. Nagataki
\inst{4}\fnmsep\inst{5}
\and
G. Ferrand
\inst{4}\fnmsep\inst{5}
\and
E. Greco
\inst{1}\fnmsep\inst{3}
\and
G. Peres
\inst{1}\fnmsep\inst{3}
\and
D.C. Warren
\inst{5}
\and
F. Bocchino
\inst{3}
}

\institute{Università degli studi di Palermo, Dipartimento di Fisica e Chimica, Via Archirafi 36 - 90123 Palermo, Italy
\and
INAF/IASF Palermo, Via Ugo La Malfa 153, I-90146 Palermo, Italy
\and
INAF-Osservatorio Astronomico di Palermo, Piazza del Parlamento 1, 90134 Palermo, Italy
\and
Astrophysical Big Bang Laboratory, RIKEN Cluster for Pioneering Research, 2-1 Hirosawa, Wako, Saitama 351-0198, Japan
\and
Interdisciplinary Theoretical \& Mathematical Science Program (iTHEMS), RIKEN, 2-1 Hirosawa, Wako, Saitama 351-0198, Japan
}

\date{Received ; accepted }

\abstract
{The manifold phases in the evolution of a core-collapse (CC) supernova (SN) play an important role in determining the physical properties and morphology of the resulting supernova remnant (SNR). Thus, the complex morphology of SNRs is expected to reflect possible asymmetries and structures developed during and soon after the SN explosion.}
{The aim of this work is to bridge the gap between CC SNe and their remnants by investigating how post-explosion anisotropies in the ejecta influence the structure and chemical properties of the remnant at later times.}
{We performed three-dimensional magneto-hydrodynamical simulations starting soon after the SN event and following the evolution of the system in the circumstellar medium (CSM) (consisting of the wind of the stellar progenitor), for 5000 years, obtaining the physical scenario of a SNR. Here we focused the analysis on the case of a progenitor red supergiant of $19.8$ M$_{\odot}$. We also investigated how a post-explosion large-scale anisotropy in the SN affects the ejecta distribution and the matter mixing of heavy elements in the remnant, during the first 5000 years of evolution.}
{In the case of a spherically symmetric SN explosion without large-scale anisotropies, the remnant roughly keeps memory of the original onion-like layering of ejecta soon after the SN event. Nevertheless, as the reverse shock hits the ejecta, the element distribution departs from a homologous expansion, because of the slowing down of the outermost ejecta layers due to interaction with the reverse shock. In the case of a large-scale anisotropy developed after the SN, we found that the chemical stratification in the ejecta can be strongly modified and the original onion-like layering is not preserved. The anisotropy may cause spatial inversion of ejecta layers, for instance leading to Fe/Si-rich ejecta outside the O shell, and may determine the formation of Fe/Si-rich jet-like features that may protrude the remnant outline. The level of matter mixing and the properties of the jet-like feature are sensitive to the initial physical (density and velocity) and geometrical (size and position) initial characteristics of the anisotropy.}
{}


\maketitle

\section{Introduction}

Core-collapse (CC) supernovae (SNe), the final fate of massive stars, are known to play a major role in the dynamical and chemical evolution of galaxies (by injecting mass and energy) and in driving the chemical enrichment of the diffuse gas. Nevertheless, many aspects of the processes governing the SN engine and the final stages in the evolution of their progenitor stars are not fully understood. Severe limitations in these studies are the rarity of SN events in our galaxy (on average about one every 50 years; ~\citealt{Diehl_2006}), the large distances of extragalactic SNe which make them unresolved point-like sources, and their unpredictability and transient nature. These issues make extremely challenging to extract, from observations of SNe, key information on the progenitor stars and on the explosion processes associated with SNe.

Otherwise, nearby supernova remnants (SNRs), i.e. what is left over of SN explosions, are extended sources for which the structure and chemical composition of the stellar debris ejected in the SN outburst (i.e. the ejecta) can be studied in detail. More specifically, young and nearby SNRs encode valuable information in their expanding debris and evolution, that can reveal details about the inherent asymmetries of the explosion itself and the products of nucleosynthesis occurred during the explosion. Moreover, SNRs probe the circumstellar medium (CSM) surrounding the SNe, which can be shaped by the progenitor stars through their powerful stellar winds. However, with increasing age, all these pieces of information, which are useful for establishing a SN-SNR connection, are diluted by the remnant’s increasing interaction with the surrounding environment. Thus  young and nearby remnants are the most likely candidates to investigate the intimate link that exists between the morphological properties of a SNR and the complex phases in the SN explosion, although in the Milky Way less than a dozen SNRs are $<$ 1000 years old~\citep{Ferrand_2012, Green_2014}.

Hydrodynamic/magnetohydrodynamic (HD/MHD) models can be powerful tools to bridge the gap between CC-SNe and their remnants. However, up to recent years, the key strategy was to describe either the SN evolution or the expansion of the remnant due to the very different time and space scales involved in these two phases of evolution and the inherent three-dimensional (3D) nature of the phenomenon. As a result, SN models described the early SN evolution, leaving out an accurate description of its subsequent interaction with the ambient environment; SNR models described only the expansion of the remnant and its interaction with the ambient medium, leaving out a self-consistent and accurate description of the distribution of mass and energy of the ejecta soon after the SN explosion. Only recently, seminal studies (\citealt{Patnaude_2015, Utrobin_2015, Orlando_2015, Orlando_2016, Wongwathanarat_2017, Orlando_2019, Ferrand_2019, Orlando_2020}) have described the whole evolution from the SN to the SNR and have shown that HD/MHD models can be very effective in studying the SN-SNR connection. Thus these studies bring together two distinct physical scenarios (SN and SNR) on the logical and temporal level, creating continuity between the two.

Up to date, the mixing of chemically homogeneous layers of ejecta after the SN event has been investigated in detail only in the aftermath of SN explosions (e.g. \citealt{Kifonidis_2006, Joggerst_2009, Joggerst_2010}), not considering the transition from the phase of SN to that of SNR. Very recently, \cite{Ono_2020} studied the matter mixing in aspherical CC-SNe through accurate three-dimensional simulations with the aim to link the ejecta structure soon after the shock breakout at the stellar surface (i.e. when the shock produced by the explosion leaves the stellar surface) with the nature of the progenitor star. 

Here, we aim at analyzing how matter mixing occurs at later times during the remnant expansion and interaction with the CSM to investigate how the various chemically homogeneous layers at the time of the explosion map into the resulting abundance pattern of the SNR, and how large-scale anisotropies typically observed in CC-SNe (possibly originated by instabilities developed during the CC; e.g.~\citealt{Nagataki_1997, Nagataki_1998, Nagataki_2000, Kifonidis_2006, Takiwaki_2009, Gawryszczak_2010, Wongwathanarat_2017, Bear_2018}) affect the evolution of these layers. To this end, we developed a 3D MHD model describing the evolution of the remnant from the onset of the SN to the development of its remnant at the age of 5000 years, focusing on the case of a progenitor red supergiant (RSG). We analyzed the effects of large-scale anisotropies that may develop soon after the SN on the final remnant morphology. These anisotropies may be responsible for the knotty and clumpy ejecta structure observed at different wavelengths in many CC SNRs (e.g. the Vela SNR, \citealt{Aschenbach_1995, Miceli_2008, Garcia_2017}; G292.0+1.8, \citealt{Park_2004}; Puppis A, \citealt{Katsuda_2008}; and Cassiopeia A, \citealt{Fesen_2006, Milisavljevic_2013}; N132D, \citealt{Law_2020}) which is commonly interpreted as a product of HD instabilities developed during the complex phases of the SN explosion \citep{Gawryszczak_2010}. However, it is still unclear the role of these anisotropies in the evolution of the chemical stratification of the remnant.

The paper is organized as follows: in Sect.~\ref{method} we describe the MHD model, in Sect.~\ref{Results} we discuss the results of the simulations, and in Sect.~\ref{Conclusions} we draw our conclusions.

\section{The numerical setup}\label{method}

The model describes the evolution of a SNR from soon after the CC of the progenitor star to the full-fledged remnant at the age of 5000 years, following the expansion of the remnant through its pre-SN environment. As in~\cite{Orlando_2016}, the initial distribution of ejecta is derived from a 1D SN model which describes the SN evolution from the CC to the shock breakout at the stellar surface, covering about 24 hours of evolution. The output of the SN simulation provided the initial radial profile of the ejecta distribution, including their chemical composition. Then we mapped this output into the 3D domain and started the 3D SNR simulations. 

\subsection{The initial conditions}
\label{ini_cond}
As initial conditions for our 3D SNR simulations we adopted one of the SN simulations presented in \cite{Ono_2020}. These authors performed both 1D and 3D HD simulations of SN explosions with the aim of investigating the matter mixing during the expansion of the blast wave through the stellar interior. Their 1D simulations assume spherical symmetry, whereas the 3D simulations consider aspherical explosions, triggered by an initial asymmetric injection of kinetic and thermal energy around the composition interface of the iron core and silicon-rich layer. The simulations were initialized by considering different pre-SN stellar models, including progenitor blue and red supergiants, and were performed with the adaptive-mesh refinement HD/MHD code FLASH~\citep{Fryxell_2000}. 

The SN model takes into account the effects of gravity (both self-gravity and gravitational effects of the central proto-neutron star), the fallback of material on the central compact object, the explosive nucleosynthesis through a nuclear reaction network including 19 species (n, p, $^1$H, $^3$He, $^4$He, $^{12}$C, $^{14}$N, $^{16}$O, $^{20}$Ne, $^{24}$Mg, $^{28}$Si, $^{32}$S, $^{36}$Ar, $^{40}$Ca, $^{44}$Ti, $^{48}$Cr, $^{52}$Fe, $^{54}$Fe and $^{56}$Ni), the feedback of nuclear energy generation, and the energy deposition due to radioactive decays of isotopes synthesized in the explosion (see \citealt{Ono_2020} for more details).
For our purposes, we considered the case of a 1D SN simulation from a non-rotating, solar metallicity progenitor RSG with a main sequence mass 19.8~M$_{\odot}$ which reduces to 15.9~M$_{\odot}$ at collapse~\citep{Sukhbold_2016}, and with injected energy $2.5 \times 10^{51}$ erg (see Sect. 3.1 in \citealt{Ono_2020}).
 
 In our SNR simulations we are not aiming at describing in detail the effects of matter mixing  before the shock breakout (indeed carefully modelled and analyzed through the 3D SN simulations by~\citealt{Ono_2020}).
 The motivation of this work is to highlight the contribution to matter mixing during the expansion of the blast wave through the CSM and to investigate the stability of the 1D onion shell structure of the progenitor star during the evolution of the SNR. To this end, we explored the case of a spherically symmetric SN explosion and the case of a SN characterized by a large-scale anisotropy. In this latter case, once the output of the 1D SN simulation is mapped into the 3D domain, we parametrized the initial large-scale anisotropy of ejecta that may have been developed soon after the SN explosion. The aim was to investigate how the matter mixing and the final remnant morphology depend on the physical (density and velocity) and geometric (position and size) properties of the initial anisotropy. 
 
\begin{figure}
   \centering
   \includegraphics[width=\hsize]{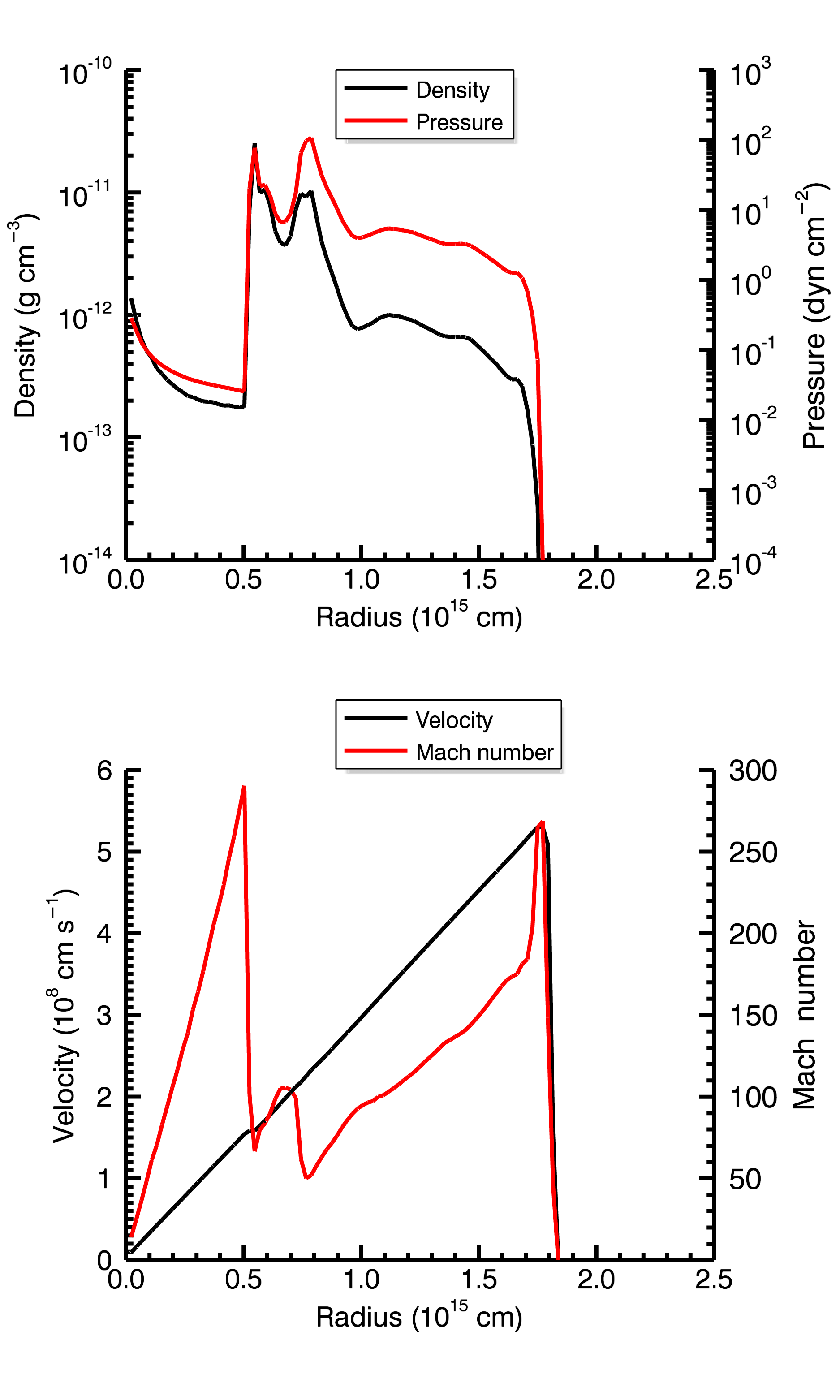}
      \caption{Initial conditions for the 3D simulations. \emph{Top panel:} radial profiles of initial ejecta density (black line) and pressure (red line). \emph{Bottom panel}: radial profiles of initial ejecta velocity (black line) and Mach number (red line).}
         \label{fig:init_plot}
   \end{figure}

\begin{figure}
   \centering
   \includegraphics[width=\hsize]{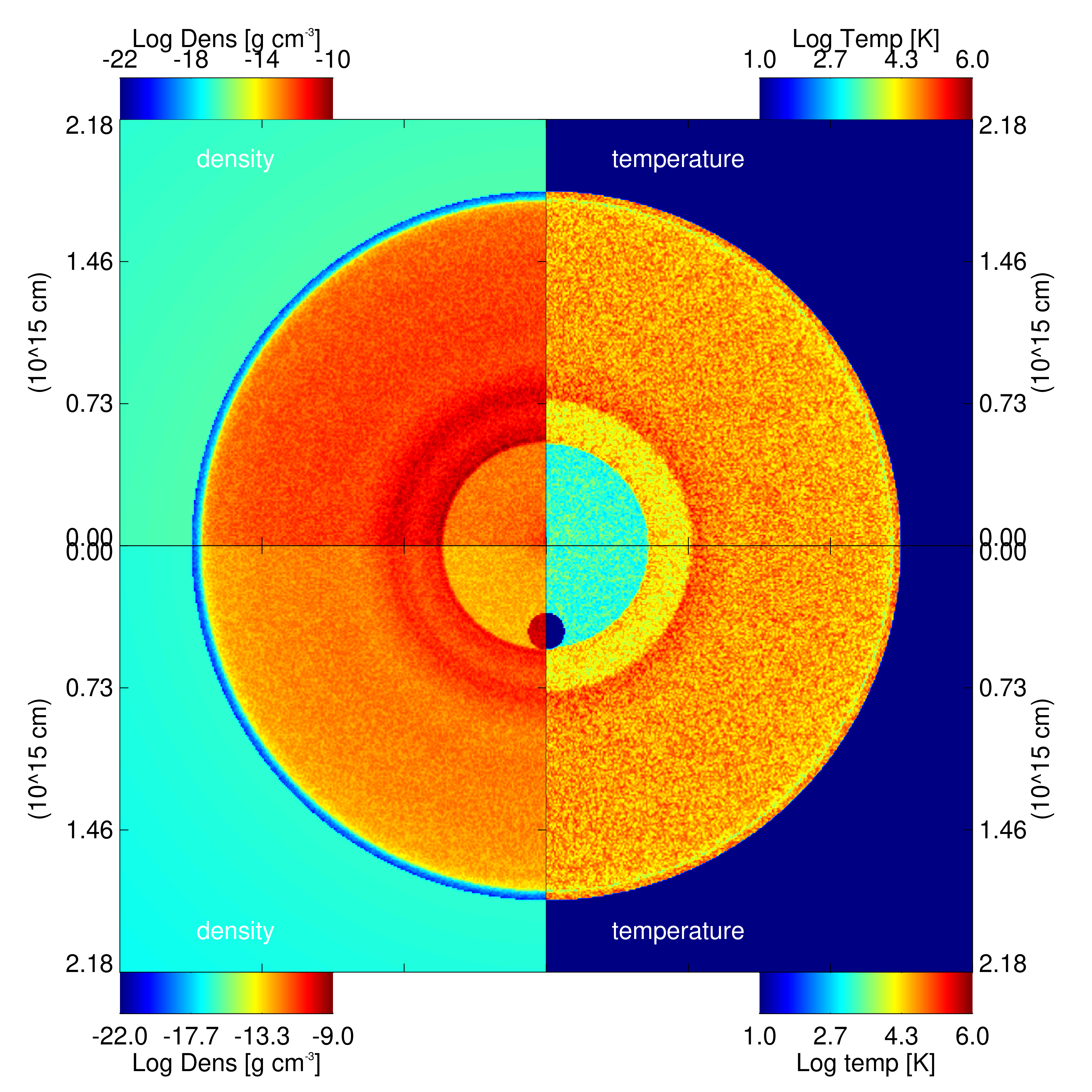}
      \caption{Density (left-hand quadrants) and temperature (right-hand quadrants) sections ($x,0,z$) showing examples of the initial conditions. The values are color-coded according to the scale shown for each quadrant. Upper quadrants show the case of a spherically symmetric explosion, lower quadrants show a case with a dense, isobaric spherical anisotropy (namely run Si-R5-D750-V5 in Table~\ref{tab:summary}).
              }
         \label{fig:init_confr}
   \end{figure}

The physical properties of the initial condition, i.e. the output from the CC-SN simulation, consist of an initial remnant at $t \sim 6\times 10^4$ s ($\approx 17$~hours) after the CC with a radius of $\sim$ 1.85$\times$10$^{15}$ cm, an ejecta mass of $14.4$ M$_{\odot}$ and a kinetic energy of $1.35 \times 10^{51}$ erg. At this epoch, hours after the breakout of the shock at the stellar surface, the ejecta expand almost homologously through the tenuous wind of the progenitor star (e.g.~\citealt{Gawryszczak_2010, Wongwathanarat_2017, Ono_2020}). The initial radial density profile of the ejecta presents a less dense inner region surrounded by a denser shell located in the region of radius $6.2 \times 10^{14}$ cm, corresponding to about $30\%$ of the radius of the ejecta (see Figs.~\ref{fig:init_plot} and~\ref{fig:init_confr}; see also \citealt{Ono_2020} for a description of the evolution of the SN and the formation of the large-scale structure in the density radial profile). This density profile arises from the SN simulation and depends on the progenitor star considered. Here we considered a generic case of a RSG; a different progenitor as, for instance, a blue supergiant could produce a density profile much smoother with structures that may significantly differ from the dense shell discussed above (see e.g. \citealt{Ono_2020}). The velocity profile (see Fig.~\ref{fig:init_plot}) increases linearly with the radius of the ejecta from 0 at the origin to a maximum of $5.3 \times 10^8$ cm s$^{-1}$ at the external shell of the initial remnant.

\subsection{Modeling the evolution of the SNR}\label{SNRmod}

The numerical setup for the SNR simulations is analogous to that described in~\citet{Orlando_2019, Orlando_2020}. Since the remnant expansion is described during the free expansion and Sedov expansion phases, the radiative losses (see~\citealt{Orlando_2005}, also treated in~\citealt{Miceli_2006, Lee_2015}) are neglected. We adopted a 3D Cartesian coordinate system ($x$, $y$, $z$) and we carried out the simulations by numerically solving the full time-dependent ideal MHD equations in conservative form:

	\begin{equation}
	\frac{\partial\rho}{\partial t} + \nabla\cdot(\rho\vec{v})=0
	\end{equation}
	
	\begin{equation}
	\frac{\partial\rho\vec{v}}{\partial t} + \nabla\cdot\left[\rho\vec{v}\vec{v} - \vec{B}\vec{B} + \vec{I} P_{\textrm{*}} \right]^T = 0
	\end{equation}
	
	\begin{equation}
	\frac{\partial\rho E}{\partial t} + \nabla\cdot[\vec{v}(\rho E + P_{\textrm{*}}) - \vec{v}\cdot \vec{B}\vec{B}] = 0
	\end{equation}
	
	\begin{equation}
	\frac{\partial B}{\partial t}+\nabla\cdot(\vec{v}\vec{B}-\vec{B}\vec{v}) = 0
	\end{equation}

\noindent
where

   \begin{align*}
   P_{\textrm{*}} = P + \frac{B^2}{2}, \qquad E = \epsilon+\frac{1}{2}v^2 +\frac{1}{2}\frac{B^2}{\rho}
	\end{align*}
	
\noindent
are the total pressure and the total gas energy (internal energy, $\epsilon$, kinetic energy, and magnetic energy), respectively, $\vec{B}$ is the magnetic field, $\rho$ is the density, $\vec{v}$ is the velocity, $\vec{I}$ is the identity matrix, $P$ is the pressure and $t$ is the time. The set of equations is closed by the equation of state (EOS) for ideal gases, $P=(\gamma - 1)\rho\epsilon$,  where $\gamma = 5/3$ is the adiabatic index.

The 3D simulations of the expanding SNR were performed using PLUTO~\citep[see][]{Mignone_2007}, a numerical code developed for the solution of hyper-sonic flows widely and successfully used in several studies. The MHD equations are solved using the MHD module available in PLUTO, configured to compute inter-cell fluxes with the Harten-Lax-van Leer discontinuities (HLLD) approximate Riemann solver, while third order in time is achieved using a Runge-Kutta scheme. To preserve the condition of $\nabla\cdot B = 0$ a mixed hyperbolic/parabolic divergence cleaning technique is used~\citep{Mignone_2007}. 

The distribution of elements is traced through a set of advection equations:

\begin{equation}
\label{eqn:tracer}
\frac{\partial \rho X_{\textrm{i}}}{\partial t}+\nabla\cdot(\rho X_{\textrm{i}} v) = 0   
\end{equation} 

\noindent
which are solved for each of the elements considered in addition to the MHD equations, where $X_{\textrm{i}}$ is the mass fraction of the element of index $i$, and $i$ runs over the 19 species listed in Section~\ref{ini_cond}. Unlike the 1D SN simulation used as initial condition, the 3D SNR simulations do not include the radioactive decays of the explosive nucleosynthesis products synthesized in the SN explosion, such as $^{56}$Ni or $^{44}$Ti. The main effect of radioactive decay would be a slight inflation of the instability-driven structures developed in regions near the decaying elements~\citep{Gabler_2017}. Since the decay time of these species is much lower than the evolution time analyzed here, we assumed at the beginning of our simulations that all the $^{56}$Ni and $^{44}$Ti are already decayed in their decaying products ($^{56}$Fe and $^{44}$Ca respectively; see \citealt{Nadyozhin_1994, Ahmad_2006}).

Our computational domain describes only one octant of the whole remnant to reduce the computational cost. At the initial condition, the explosion is located at the origin (0,0,0) of the coordinate system.  To follow the wide range of space and time scales covered during the expansion of the remnant from the onset of the explosion to the first 5000 years, we adopted the same approach as~\citet{Orlando_2019}. Initially the domain extends between 0 and $2.19 \times 10^{15}$ cm in all directions and is covered by a uniform grid with $256^3$ zones. Then the domain is gradually extended as the forward shock propagates outward and the physical values are remapped in the new domains. More specifically, each time the forward shock gets close to 90\% of the domain size, the domain is extended by a factor of 1.2 in all directions, maintaining a uniform grid of $256^3$ grid points. As discussed by~\citet{Ono_2013} this approach is reliable because it does not introduce errors larger than 0.1\% after 40 remappings. In such a way, the finest spatial resolution is $\sim10^{13}$ cm at the beginning of our 3D simulations and, because of the expansion of the system, the resolution is $\sim10^{17}$ cm at the end of simulation time. The simulations, therefore, span from an initial domain of ($2.19 \times 10^{15}$ cm)$^3$ up to ($\sim 5.8 \times 10^{19}$ cm)$^3$. Equatorial symmetric boundary conditions~\footnote{Variables are symmetrized across the boundary and normal components of vector fields flip signs, with the exception of the magnetic field in which only the transverse component flip sign.} with respect to $\textrm{x}$=0, $\textrm{y}$=0 and $\textrm{z}$=0 planes are considered, while fixed pre-shock values conditions are assumed at all other boundaries. We checked the effects of spatial resolution on the remnant evolution in Appendix A, and verified that our main conclusions are not significantly affected by the resolution adopted (see Appendix A for further details).

The ejecta distribution is expected to be characterized by small-scale clumping of material and larger-scale anisotropies (e.g. \citealt{Orlando_2016}). Thus, following~\citet{Orlando_2012}, we introduced small-scale perturbations in the initial ejecta distribution, mimicking the small-scale clumping of ejecta. These inhomogeneities were modeled as a per-cell random density perturbations derived from a power-law probability distribution. All the clumps have an initial size of $\approx 10^{13}$ cm and the density perturbation of each clump is calculated as the ratio of the mass density of the resulting clump to the local (unperturbed) average density in the region occupied by the clump; the maximum density perturbation allowed is a factor of 4. In Sect.~\ref{sec:anisotropy} we discuss how post-explosion large-scale anisotropies in the ejecta distribution are modeled.

The CSM around the progenitor star has been described as a spherically symmetric steady wind with a gas density proportional to $r^{-2}$ (where $r$ is the radial distance from the center of explosion): 

	\begin{equation}
	\rho_{\textrm{wind}} = \rho_{\textrm{ref}} \Big(\frac{r_{\textrm{ref}}}{r}\Big)^2~.
	\end{equation}

\noindent
We assumed that the wind is isobaric, with pressure	$P_{\textrm{wind}} \sim 10^{-10}$ dyn cm$^{-2}$ to prevent the wind to evolve towards the boundaries of the domain, causing "jumps" in the wind's profile whenever the domain is extended. This assumption does not affect the results because the forward shock propagates much faster than the stellar wind during the whole evolution.

The CSM is characterized by three parameters: the reference density, the radial distance of the reference density, and the temperature ($n_{\textrm{ref}} = 1$~cm$^{-3}$, $r_{\textrm{ref}} = 2.5$ pc, and $T_{\textrm{ref}} = 100$~K, respectively). The values considered are within the range of typical values for a progenitor RSG, which is subject to high mass-loss rates ($10^{-5}-10^{-4}$ M$_{\odot}$ yr$^{-1}$) and very slow winds (20 - 100 km~s$^{-1}$)~\citep{Dwarkadas_2005, Crowther_2001}. Assuming $\dot{M} = 5\times 10^{-5}$ M$_{\odot}$~yr$^{-1}$ and $v_{\textrm{w}} = 20$ km~s$^{-1}$, we derived the reference mass density as $\rho_{\textrm{ref}} = \dot{M}/(4\pi v_w r_{\textrm{ref}}^2)$ which corresponds to a particle number density $n_{\textrm{ref}} \approx 1$~cm$^{-3}$. In such a tenuous environment, the radiative losses would become relevant after $t_{rad} > 10^4$ years (see~\citealt{Blondin_1998, Orlando_2005, Miceli_2006, Lee_2015}), a timescale much greater than the simulated time ($t = 5000$ years).
	
The simulations include the effect of an ambient magnetic field. This field is not expected to influence the overall dynamics of the system which is characterized by high values of the plasma $\beta$ (i.e. the plasma pressure dominates over the magnetic one). We note that MHD codes are generally more diffusive than HD codes because they use approximate Riemann solvers. However, we decided to include the effects on an ambient magnetic field in our simulations because the field can play a relevant role in limiting the growth of HD instabilities that would develop at the border of ejecta clumps and anisotropies and that are responsible for their fragmentation~\citep{Fragile_2005, Shin_2008, Orlando_2008, Orlando_2012, Orlando_2019}. Following \cite{Orlando_2019}, the field configuration adopted is the "Parker spiral" that is the field resulting from the rotation of the  progenitor star and the expanding stellar wind \citep{Parker_1958}. The Parker spiral can be described in spherical coordinates ($r$, $\theta$, $\phi$) as:

\begin{align}
	B_{\textrm{r}}      &= \frac{A_{\textrm{1}}}{r^2} \\
	B_{\phi} &=-\frac{A_{\textrm{2}}}{r}\sin\theta.
\noindent
\end{align}
The previous parameters are set to $A_{\textrm{1}} = 3 \times 10^{28}$ G cm$^2$ and $A_{\textrm{2}} = 8 \times 10^{10}$ G cm in order to produce a magnetic field of the order of $10$ G at the surface of the progenitor star (e.g. \citealt{Tessore_2017} and references therein).

\subsection{Post explosion large-scale anisotropy}
\label{sec:anisotropy}

 We considered the case of a post-explosion large-scale anisotropy located in the inner initial ejecta distribution. Our aim is to study how/whether the remnant structure and morphology keep memory of post-explosion anisotropies and if these anisotropies have an effect on the evolution and mixing of the different metal-rich layers of the progenitor star. The anisotropy included in our model does not arise from the 1D SN simulation, but is set in the initial conditions of our SNR model. In fact, explaining the physical origin of these initial anisotropies is well beyond the scope of this work; indeed we aimed at studying the effects of a large scale post-explosion anisotropy on the evolution of a SNR and at investigating how the different initial parameters of the anisotropy might determine the final shape of the remnant.

 The anisotropy is described as an overdense sphere in pressure equilibrium with the surrounding ejecta, whose center lies on the z-axis (therefore only a quarter of the sphere is modelled). We investigated possible numerical effects which may develop by assuming the overdense sphere propagating along the $z$-axis, by exploring a case of a clump with its center located at $\sim 45^{\circ}$ in the plane $(x,0,z)$ (see Appendix B for further details). We found that simulations assuming the sphere propagating either along the $z$-axis or at $\sim 45^{\circ}$ in the plane $(x,0,z)$ produce very similar results on the time-scale considered (5000~years) and we concluded that the numerical effects (if any) due to the propagation of the bullet along the $z$-axis have a negligible impact on the evolution of the ejecta bullet.
 
 The geometrical properties of the anisotropy are parametrized by its distance from the center ($D$), and its radius ($r$). We considered two values of $D$, associated with different layers of the exploding star, namely: the interface between $^{56}$Fe/$^{28}$Si-rich regions at $D = 0.24$~$R_{\textrm{SNR}}$ (where $R_{\textrm{SNR}}$ is the initial radius of remnant) and the $^{56}$Fe-rich region  at $D = 0.15$~$R_{\textrm{SNR}}$. We explored two values for the clump radius, namely $r = 0.05\,R_{\textrm{SNR}}$ and $r = 0.04\,R_{\textrm{SNR}}$, both consistent with the characteristic size of clumps generated by Rayleigh-Taylor (RT) instability (seeded by flow structures resulting from neutrino-driven convection) in SN explosion simulations (e.g. \citealt{Kifonidis_2006, Gawryszczak_2010}). We explored anisotropies with density between $500$ and $750$ times that of the surrounding ejecta at distance $D$ (density contrast $\chi_{\textrm{n}}$), and with radial velocity between $2$ and $7$ times that of the surrounding ejecta (velocity contrast $\chi_{\textrm{v}}$), making sure that the maximum speed of the clump does not exceed the velocity of the fastest ejecta in the outermost layers. \cite{Wongwathanarat_2017} have shown that fast and high-density large-scale anisotropies/plumes with high concentration of $^{56}$Ni and $^{44}$Ti can emerge from neutrino-driven SN explosions. Some other authors have predicted that high-velocity ejecta components might be produced in the presence of magnetic fields that are anchored to a rapidly-rotating proto-neutron star (e.g.~\citealt{Takiwaki_2009}). The anisotropies parametrized in this paper assume the same chemical composition as the surrounding ejecta region, i.e. iron or iron/silicon when $D=0.15$ and $D=0.24$. A summary of the cases explored is given in Table~\ref{tab:summary}.

Previous simulations of ejecta bullets in a SNR, performed by \citet{Miceli_2013} and \citet{Tsebrenko_2015}, showed that clumps with initial values of $\chi_{\textrm{n}} \leq 100$ are able to reach the SNR shock front without being dispersed by the interaction with the reverse shock. At odds with those simulations which adopt power-law density profiles for the ejecta distribution and start tens of years after the SN event, our simulations include a self-consistent description of the initial ejecta density profile (derived from 1D SN simulations) which shows a high-density thick shell in the innermost regions of the ejecta (see left boxes of Fig.~\ref{fig:init_confr}). As a result, we found that higher density contrasts ($\chi_{\textrm{n}}>500$) are necessary for the clumps to overcome this overdense structure and to eventually protrude the forward shock. The maximum initial mass of the anisotropies considered is $\sim 0.25$ M$_\odot$ (< 2\% of the initial ejecta mass) with a maximum kinetic energy of $\sim 4 \times 10^{49}$ erg (< 3\% of the total initial kinetic energy of the ejecta).We note that these values are very similar to those estimated for the Si-rich jet-like feature observed in Cassiopeia A~\citep{Orlando_2016} and to those inferred for the shrapnel G in the Vela SNR~\citep{Garcia_2017}.

\begin{table}
\caption{Parameters describing the initial conditions for the models following the evolution of a post-explosion anisotropy of ejecta.}
\label{tab:summary}      
\centering          
\begin{tabular}{ c c c c c}     
\hline\hline 
\multirow{2}{*}{Model} &     $D$    &    $r$    & $\chi_{\textrm{n}}$ & $\chi_{\textrm{v}}$ \\
&  ($R_{\textrm{SNR}}$) & ($R_{\textrm{SNR}}$) &          &          \\
\hline
Si-R5-D750-V5      &   $0.24$   &  $0.05$   &  $750$   &   $5$    \\
Si-R5-D750-V4      &   $0.24$   &  $0.05$   &  $750$   &   $4$    \\
Si-R5-D750-V3      &   $0.24$   &  $0.05$   &  $750$   &   $3$    \\
Si-R5-D750-V1      &   $0.24$   &  $0.05$   &  $750$   &   $1$    \\
Si-R5-D600-V5      &   $0.24$   &  $0.05$   &  $600$   &   $5$    \\
Si-R5-D600-V4      &   $0.24$   &  $0.05$   &  $600$   &   $4$    \\
Si-R5-D600-V3      &   $0.24$   &  $0.05$   &  $600$   &   $3$    \\
Si-R5-D500-V5      &   $0.24$   &  $0.05$   &  $500$   &   $5$    \\
Si-R5-D500-V4      &   $0.24$   &  $0.05$   &  $500$   &   $4$    \\
Si-R5-D500-V3      &   $0.24$   &  $0.05$   &  $500$   &   $3$    \\
Fe-R5-D750-V7      &   $0.15$   &  $0.05$   &  $750$   &   $7$    \\
Fe-R5-D750-V6      &   $0.15$   &  $0.05$   &  $750$   &   $6$    \\
Fe-R5-D750-V5      &   $0.15$   &  $0.05$   &  $750$   &   $5$    \\
Fe-R5-D750-V3      &   $0.15$   &  $0.05$   &  $750$   &   $3$    \\
Fe-R5-D750-V2      &   $0.15$   &  $0.05$   &  $750$   &   $2$    \\
Fe-R5-D600-V7      &   $0.15$   &  $0.05$   &  $600$   &   $7$    \\
Fe-R5-D600-V6      &   $0.15$   &  $0.05$   &  $600$   &   $6$    \\
Fe-R5-D500-V7      &   $0.15$   &  $0.05$   &  $500$   &   $7$    \\
Fe-R5-D500-V6      &   $0.15$   &  $0.05$   &  $500$   &   $6$    \\
Fe-R5-D500-V5      &   $0.15$   &  $0.05$   &  $500$   &   $5$    \\
Fe-R5-D500-V3      &   $0.15$   &  $0.05$   &  $500$   &   $3$    \\
Fe-R5-D500-V2      &   $0.15$   &  $0.05$   &  $500$   &   $2$    \\
Fe-R4-D750-V7      &   $0.15$   &  $0.04$   &  $750$   &   $7$    \\
Fe-R4-D750-V6      &   $0.15$   &  $0.04$   &  $750$   &   $6$    \\
Fe-R4-D500-V7      &   $0.15$   &  $0.04$   &  $500$   &   $7$    \\
Fe-R4-D500-V6      &   $0.15$   &  $0.04$   &  $500$   &   $6$    \\

\hline
	
\end{tabular}
\end{table}
	
\section{Results}\label{Results}

\subsection{The case of a spherically symmetric explosion}\label{subsec:spher}
We first considered the case of a spherically symmetric SN explosion with small-scale (isotropically distributed) clumps  of ejecta, without including any large-scale anisotropies (hereafter, the spherical model) to study the spherically symmetric expansion of the remnant and the evolution of its chemically homogeneous layers.

\begin{figure*}
   \centering
   {\includegraphics[width=.48\textwidth]{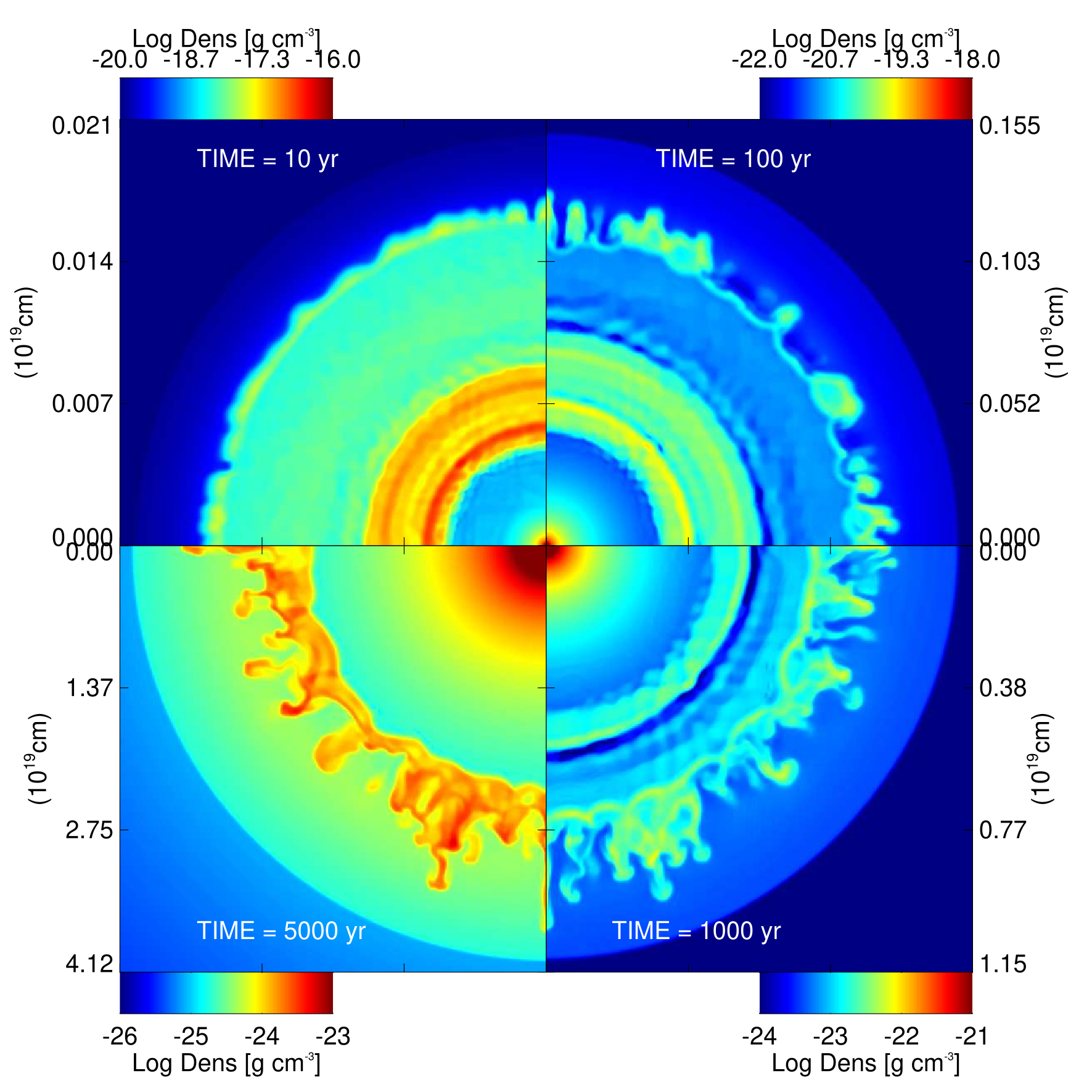}} \quad
   {\includegraphics[width=.48\textwidth]{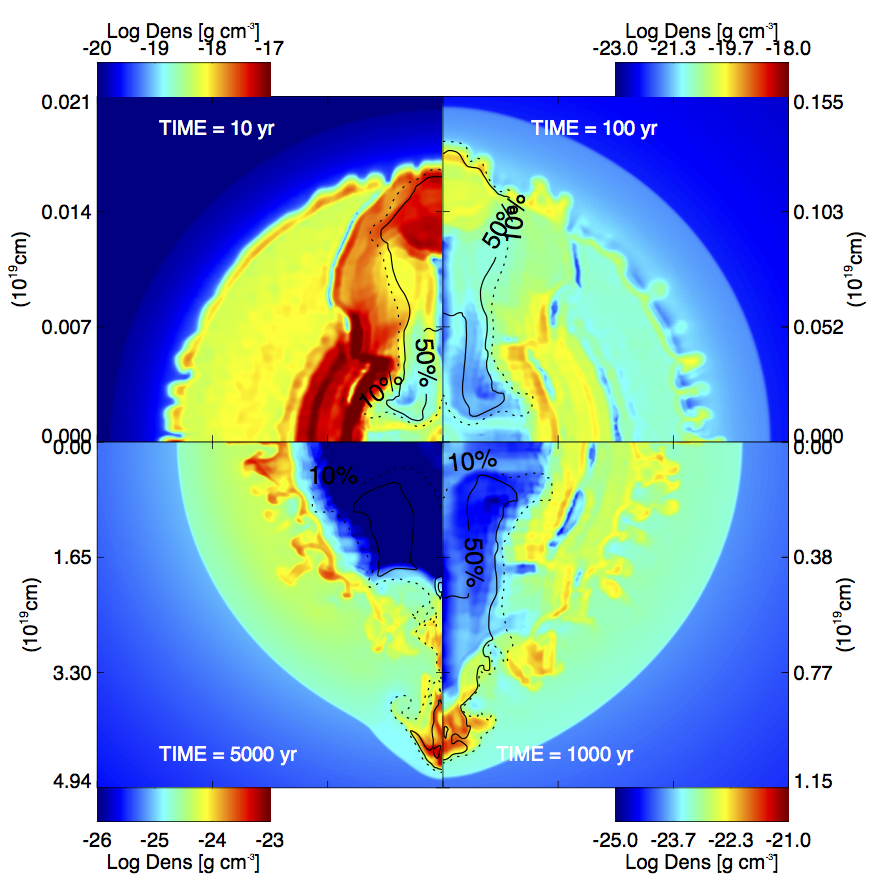}} \quad
      \caption{Density distributions in the ($x,0,z$) plane at different simulation times for a spherically symmetric explosion (left-panel) and for model Fe-R5-D750-V6 of Table~\ref{tab:summary} (right-panel). From top left clockwise: $t = 10$, $t = 100$, $t = 1000$ and $t = 5000$ years from the explosion. The units in the color bars are g cm$^{-3}$ logarithmically scaled. The colour coded density scale is shown close to each quadrant. Note the different scales used in each quadrant for both the density and the distance along the axis. In the right panel the contours enclose the computational cells consisting of the original anisotropy material by more than 50$\%$ (solid line) and 10$\%$ (dotted line). See also online Movie 1 and 2 for an animated version.}
         \label{fig:spher_dens}
   \end{figure*}

Fig.~\ref{fig:spher_dens} (see also online Movie 1 and 2) shows the 2D cross-sections through the ($x,0,z$) plane of density at different evolutionary stages for the spherical model (left panel). After $100$ years of evolution, it is possible to see HD instabilities (Rayleigh-Taylor, Kelvin-Helmholtz shear instability; e.g. \citealt{Gull_1973, Chevalier_1976, Fryxell_1991, Chevalier_1992}) developing as finger-like structures at the interface between ejecta and shocked CSM (as already shown by numerical simulations in~\citealt{Dwarkadas_1998, Wang_2001}), thus driving the mixing between the ejecta and the ambient. During this evolutionary phase, the over-dense internal shell in the ejecta expands and it is clearly visible in the density distribution. At the age of $1000$ years the reverse shock reaches the over-dense internal shell. This allows the over-dense shell to enter in the mixing-region causing the destruction of the shell itself by the HD instabilities at $t = 5000$ years (see online Movie 1). At the end of the simulation, the SNR reaches a radius of $\approx 4\times 10^{19}$ cm ($\approx 13$~pc).

Among all the elements considered in the simulations, radioactive nuclei (such as $^{56}$Ni and $^{44}$Ti), synthesized during the explosion, and their decay products are very important because they are produced in the innermost regions of the exploding star, which makes them excellent probes of the internal conditions that lead to the determination of the explosion mechanism and of the shock-wave dynamics during the earliest phases of SN outbursts. For this reason, the study of these elements is a key to connect explosion models to observations and to deduce important constraints on the underlying processes. Radioactive $^{56}$Ni and $^{44}$Ti transform into the stable isotopes $^{56}$Fe and in $^{44}$Ca, respectively, with a half-life of $<100$ years \citep{Nadyozhin_1994, Ahmad_2006}. As mentioned in Sect.~\ref{method}, since we are interested to investigate the matter mixing and evolution of the remnant at epochs longer than 100 years, and recalling that our model does not include the radioactive decay of the elements (see Section~\ref{SNRmod}), we assumed that all the $^{56}$Ni and $^{44}$Ti are already decayed in $^{56}$Fe and $^{44}$Ca respectively at the beginning of our simulations.

\begin{figure*}
   \centering
            {\includegraphics[width=.45\textwidth]{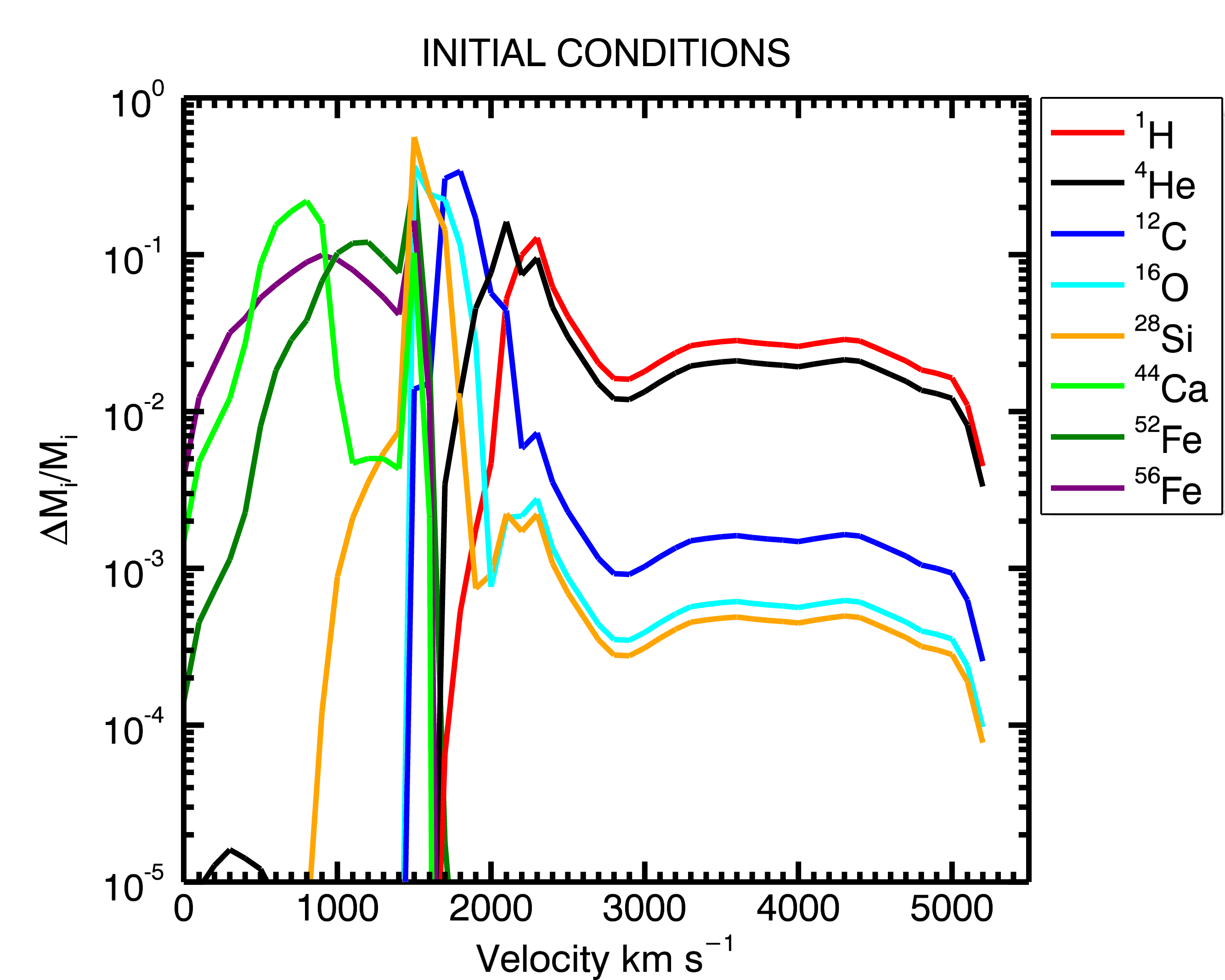}} \quad
            {\includegraphics[width=.45\textwidth]{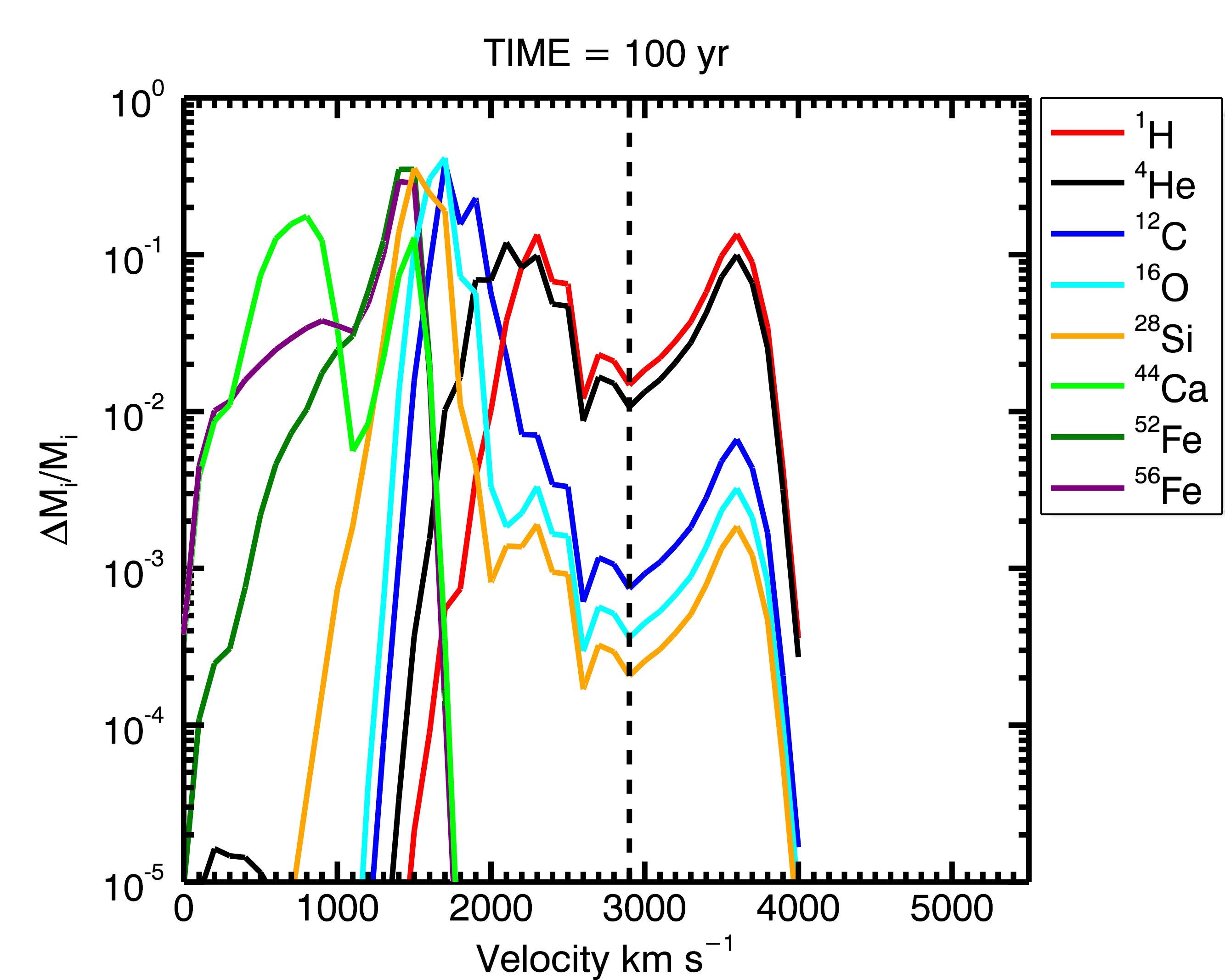}} \\
            {\includegraphics[width=.45\textwidth]{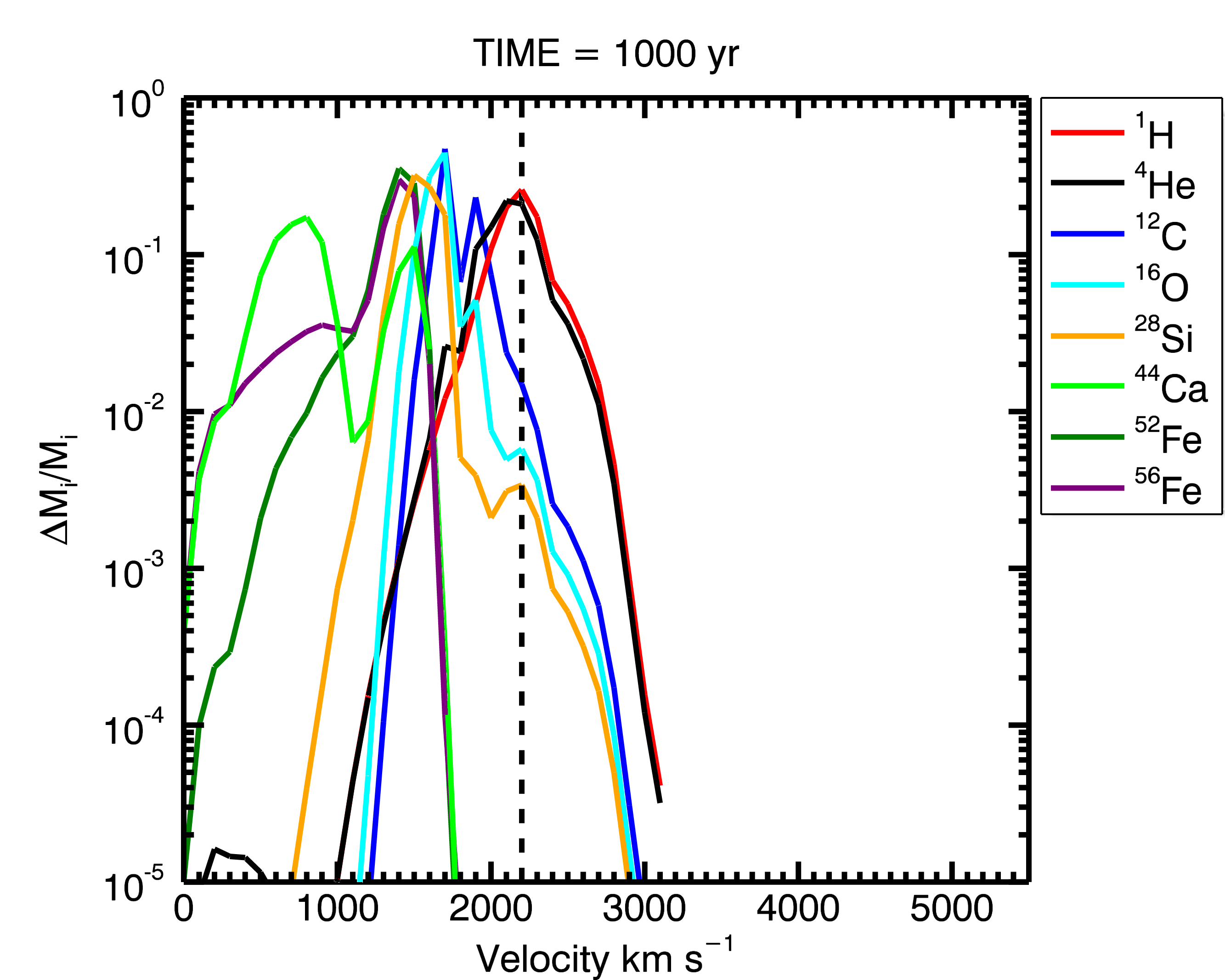}} \quad
            {\includegraphics[width=.45\textwidth]{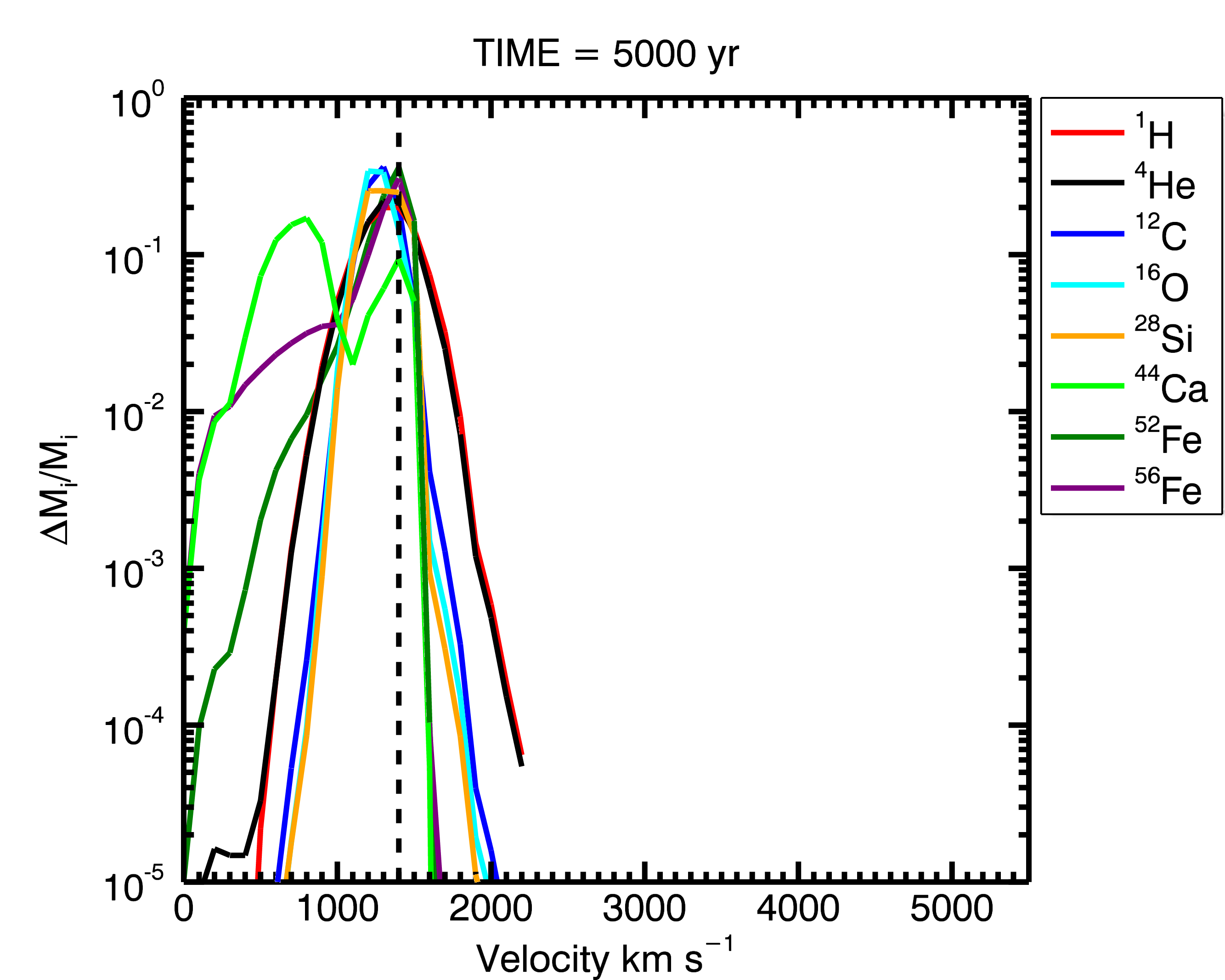}} \quad
   \caption{Mass distributions of selected elements as a function of radial velocity for the spherically symmetric explosion at the labeled years from the explosion. The upper left panel shows the initial conditions from~\cite{Ono_2020}. Only the dominant fractions are considered. M$_i$ is the total ejecta mass of element $i$, $\Delta$M$_i$ is the mass of the $i$-th element in the velocity range between $v$ and $v + \delta v$. The size of the velocity bins $\delta v$ is $100$ km s$^{-1}$. The black dashed vertical line shows the approximate reverse shock position in each panel.}
   \label{fig:spher_massfrac}
   \end{figure*}

Figure~\ref{fig:spher_massfrac} displays the ejecta mass distribution of selected elements versus the radial velocity, at different epochs during the remnant expansion. In the figure, $\Delta M_i$ is the mass of the $i$-th element in the velocity range between $v$ and  $v + \Delta v$, and the velocity is binned with $\Delta v = 100$ km s$^{-1}$.
The mass distribution in velocity space of $^{56}$Fe, $^{44}$Ca, $^{28}$Si, $^{16}$O and $^{12}$C, at the beginning of the simulations, are shown in the top left panel of Fig.~\ref{fig:spher_massfrac}. The elements appear stratified: $^{56}$Fe is concentrated in the slower and innermost part of the ejecta, approximately at $30\%$ of the radius of the initial remnant, inside the dense shell seen in the density distribution in Fig.~\ref{fig:init_confr}, along with the other heavy elements (as $^{44}$Ca). $^{28}$Si envelopes the inner $^{56}$Fe, whereas $^{16}$O and $^{12}$C are prominent at the bottom of the $^4$He shell. The outer half of the radial ejecta distribution is dominated by $^4$He and $^1$H. 

Looking at the mass distribution of the selected elements at various times we observe how the overall composition of the ejecta changes during the remnant expansion (see Fig.~\ref{fig:spher_massfrac}). The overall structure and composition of the unshocked ejecta, on the time-scale considered, do not change significantly, although some small differences in the innermost layers are present due to some numerical diffusivity of the code. This indicates that the unshocked ejecta expand almost homologously until they start to interact with the reverse shock. This is shown in Fig.~\ref{fig:spher_massfrac}, where the profiles of the elements with velocity $< 1500$ km s$^{-1}$ stay almost unchanged at all epochs. In the unshocked ejecta, we observe a large fraction of $^{56}$Fe and $^{44}$Ca, together with the low-velocity tail of the $^{28}$Si distribution. Instead, the profiles of the elements at a greater distance from the centre of the explosion, i.e. the elements with high-velocity components, are strongly influenced by their interaction with the reverse shock and the CSM. These elements are strongly decelerated in the intershock region and consequently mixed by HD instabilities. This produces a homogenization of the different profiles in these regions: the overall shapes of the high-velocity tails of the mass distributions of shocked elements are more similar to each other and characterized by steep slopes, suggesting a significant mixing between layers of different chemical composition due to HD instabilities developing at the contact discontinuity (see also \citealt{Wongwathanarat_2017} for the effects of HD instabilities on the mass distribution versus the radial velocity). 

Nevertheless, it is important to note that the chemical distribution generated during the immediate aftermath of the SN (i.e. the onion-skin chemical layering) is roughly preserved after $5000$ years of evolution. Thus, for a spherically symmetric SN explosion, it is not possible to reproduce the inversion among the innermost chemical layers (e.g. \citealt{Wang_2002, Wang_2011, Orlando_2016}) as observed, for instance, in Cassiopeia A (e.g. \citealt{Hughes_2000}), if the ejecta are characterized only by small-scale clumping.

      \begin{figure}
   \centering
   \includegraphics[width=\hsize]{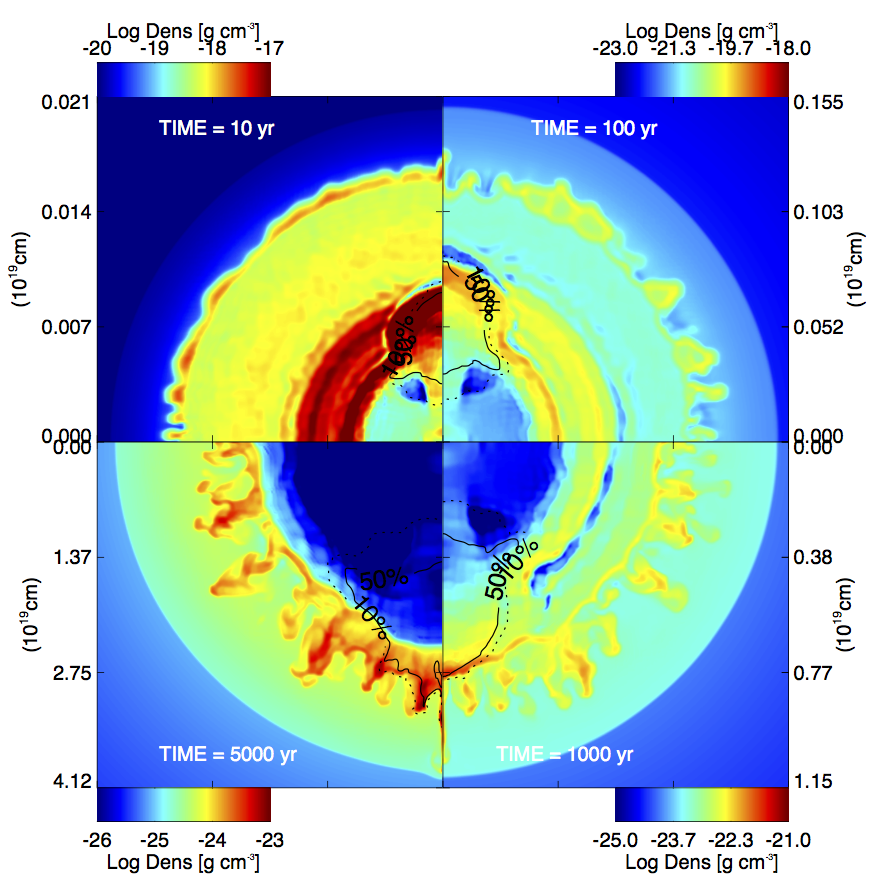}
      \caption{Same as the right-panel in Fig.~\ref{fig:spher_dens} but for model Fe-R5-D750-V3 (see Table~\ref{tab:summary})
              }
         \label{fig:clump_dens2}
   \end{figure}

\subsection{Effect of a large-scale anisotropy}
\label{sec:results_anisotropy}
As seen in Section~\ref{subsec:spher}, the ideal case of a spherically symmetric explosion cannot reproduce the complex morphology of an evolved SNR, in good agreement with the fact that most of the CC-SNe are believed to explode asymmetrically~\citep{Maeda_2008, Wongwathanarat_2013, Janka_2016, Janka_2017, O'Connor_2018, Burrows_2019}. The physical characteristics of the SN explosion generate asymmetric expansion and strong inhomogeneities that have a complex influence on the evolution of the SNR. Here we consider a possible example of a dynamical event that can cause strong mixing and overturning of the chemical layers in the ejecta. We studied the effects of a post-explosion anisotropy generated soon after the shock breakout. Multiwavelength observations have shown the presence of clump structures in the ejecta of CC SNRs~\citep{Aschenbach_1995, Miceli_2008, Garcia_2017, Park_2004, Katsuda_2008, Fesen_2006}. We model the anisotropy as a spherical clump of radius $r$, with its center located on the $z$ axis, at a distance $D$ from the origin, as explained in Sect. \ref{sec:anisotropy}.
 
As shown in Table~\ref{tab:summary}, we explored the case of a clump located either at $D = 0.15$ or at $D = 0.24$, corresponding to the initial $^{56}$Fe-rich internal layer and between the $^{56}$Fe- and $^{28}$Si-rich layers, respectively (see the top left panel in Fig.~\ref{fig:spher_massfrac}).

Right panel of Fig.~\ref{fig:spher_dens} shows a 2D cross section through the ($x, 0, z$) plane of density distribution of model Fe-R5-D750-V6 at four different evolution times (but qualitatively the overall evolution is similar for all the cases explored). The top left quadrant shows the system $10$ years after the beginning of the simulation, when the clump has already overcome the dense shell initially located at about $30\%$ of the radius of the ejecta and the topmost part of the clump has reached the reverse shock. At this stage, the clump has an elongated structure along the z-axis, from the central region of the remnant to the reverse shock. Only the head of the clump is still significantly overdense with respect to the surrounding ejecta. The top right quadrant shows the system at $100$ years. The clump has not changed its morphology, but its topmost part interacts with the intershock region. RT instabilities start to develop at the interface between ejecta and shocked CSM. After $1000$ years of evolution (bottom right quadrant) the part of the clump in the intershock region is partially eroded by the HD instabilities, while the passage of the clump has left a low-density-strip region in the inner part of the ejecta. This phase lasts $\approx$~2000 years, then the wake behind the ejecta clump is filled in eventually as the pressure gradient drives lateral expansion of the ejecta to fill the void. For model Fe-R5-D750-V7 and Si-R5-D600-V4 we found that at $\sim 2500$ years, the clump reaches a velocity of $\sim 3000$ km s$^{-1}$, similarly to what was found very recently for the runaway knot in N132D ($3650$ km s$^{-1}$, \citealt{Law_2020}).
The bottom left quadrant shows the clump at $5000$ years and it is possible to see the characteristic supersonic bow shock of the clump protruding beyond the SNR forward shock. At this evolutionary stage, the low-density region has been filled by the surrounding material. An exception to this trend is represented by models Fe-R5-D750-V3 (see Fig.~\ref{fig:clump_dens2}) and Fe-R5-D500-V3. In these two cases, the clump penetrates into the high-density shell without overcoming it, remaining stuck inside it. In this case, the clump expands together with the surrounding ejecta and, eventually, it is completely fragmented by the interaction with the reverse shock. 

   \begin{figure*}
   \centering
            {\includegraphics[width=.45\textwidth]{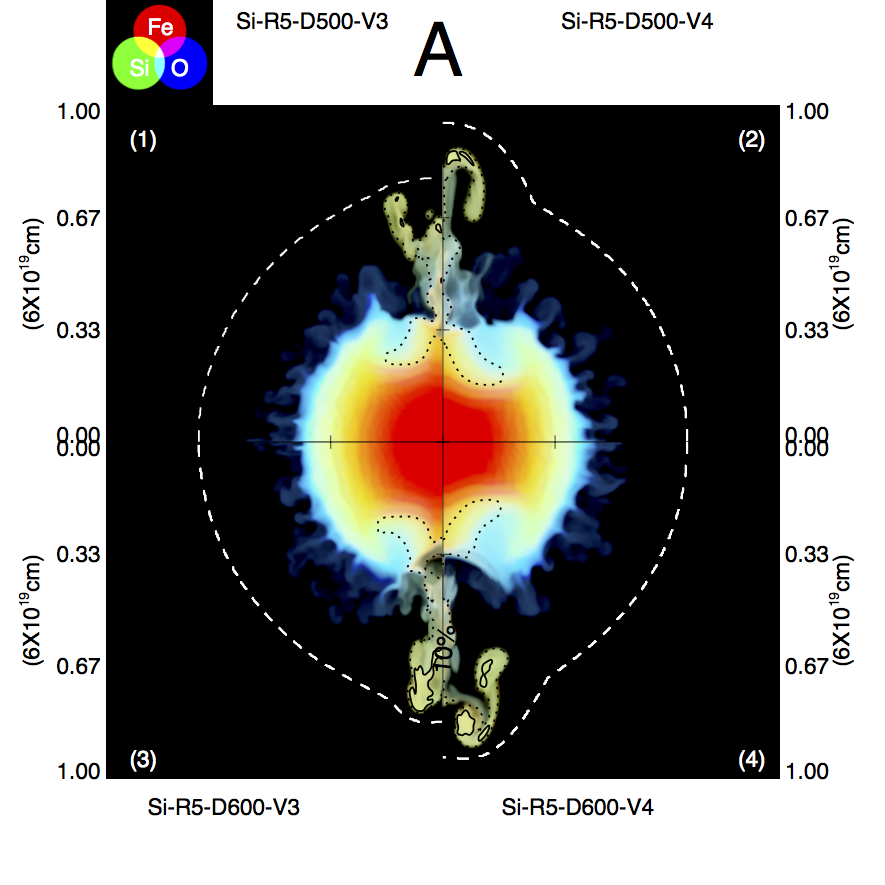}} \quad
            {\includegraphics[width=.45\textwidth]{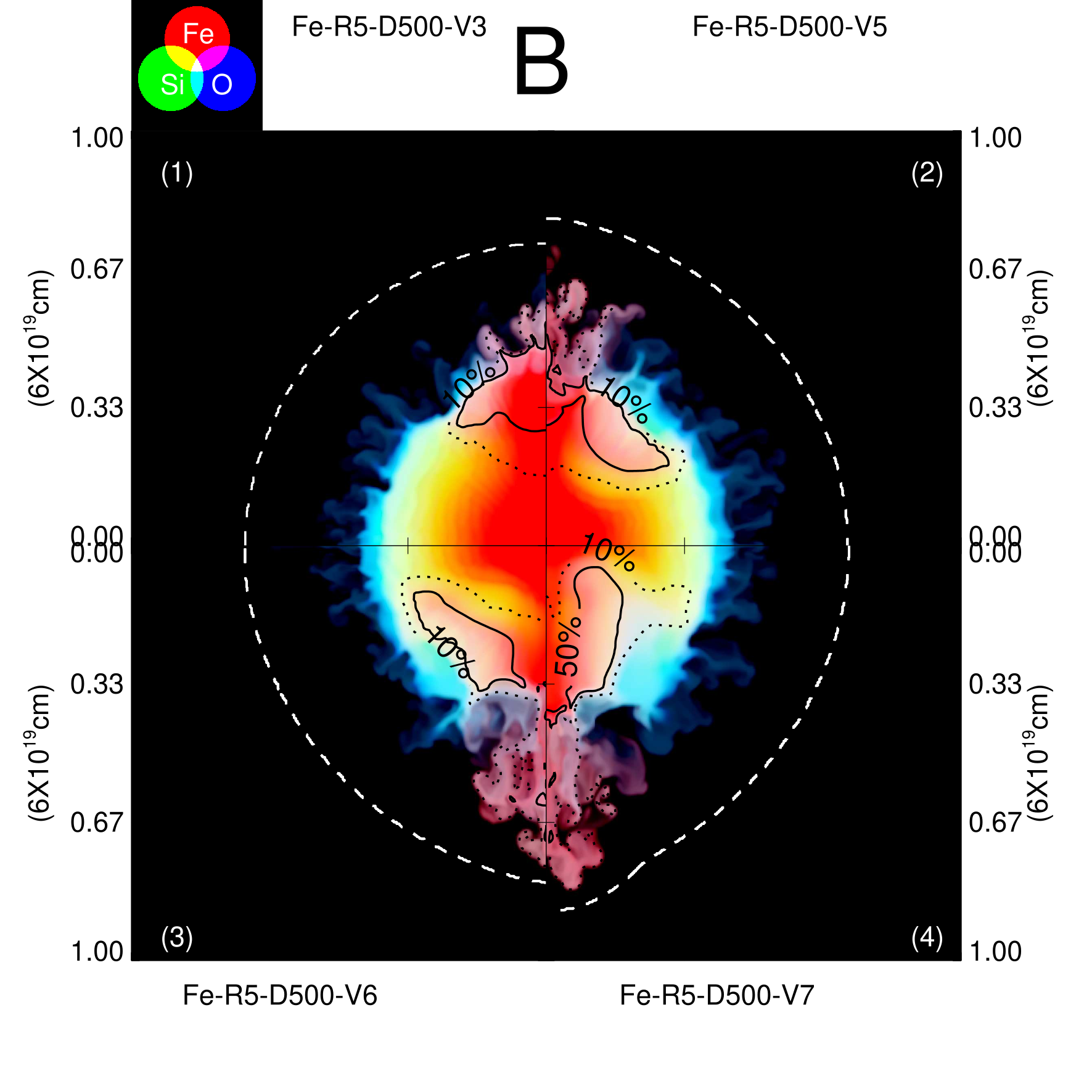}} \\
            {\includegraphics[width=.45\textwidth]{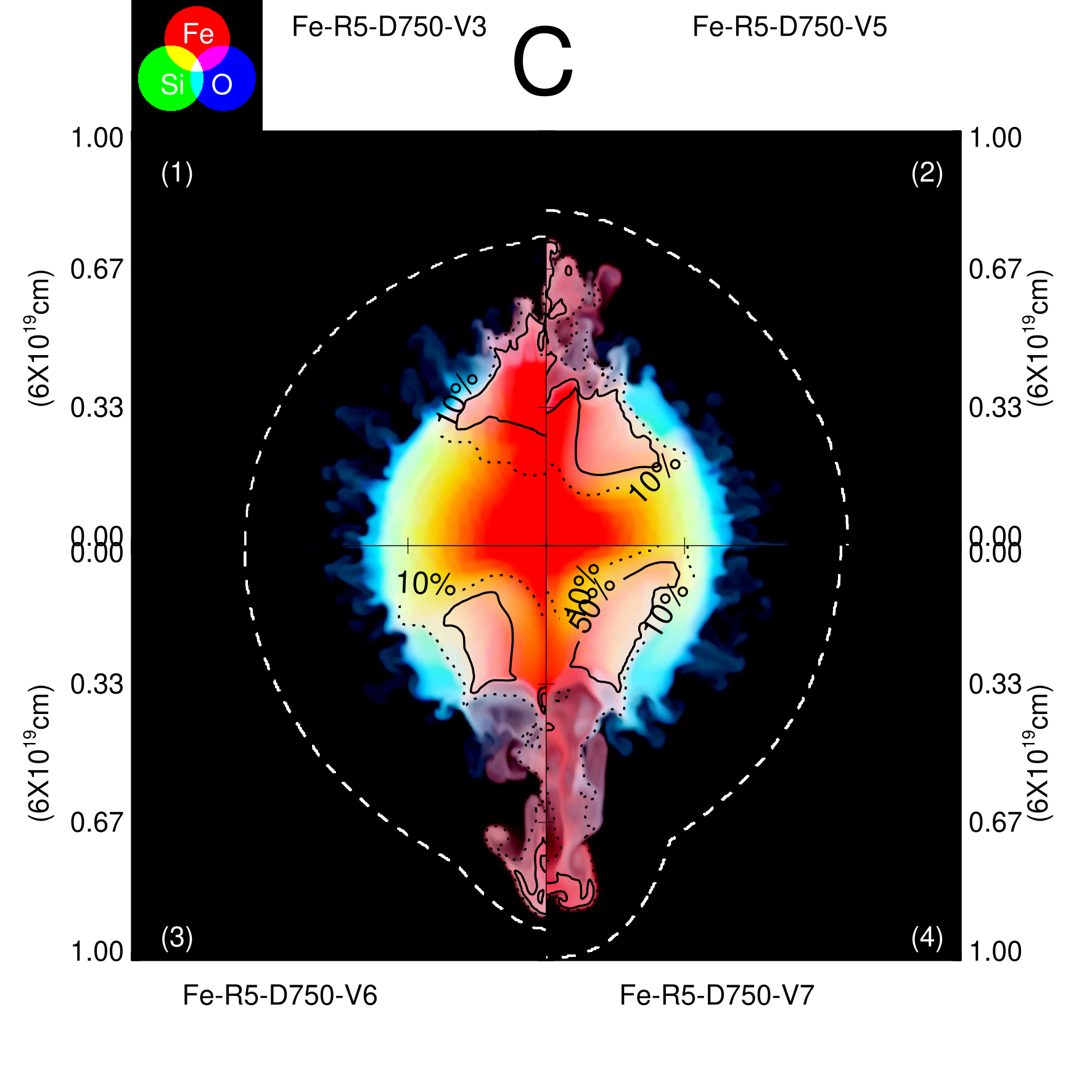}} \quad
            {\includegraphics[width=.45\textwidth]{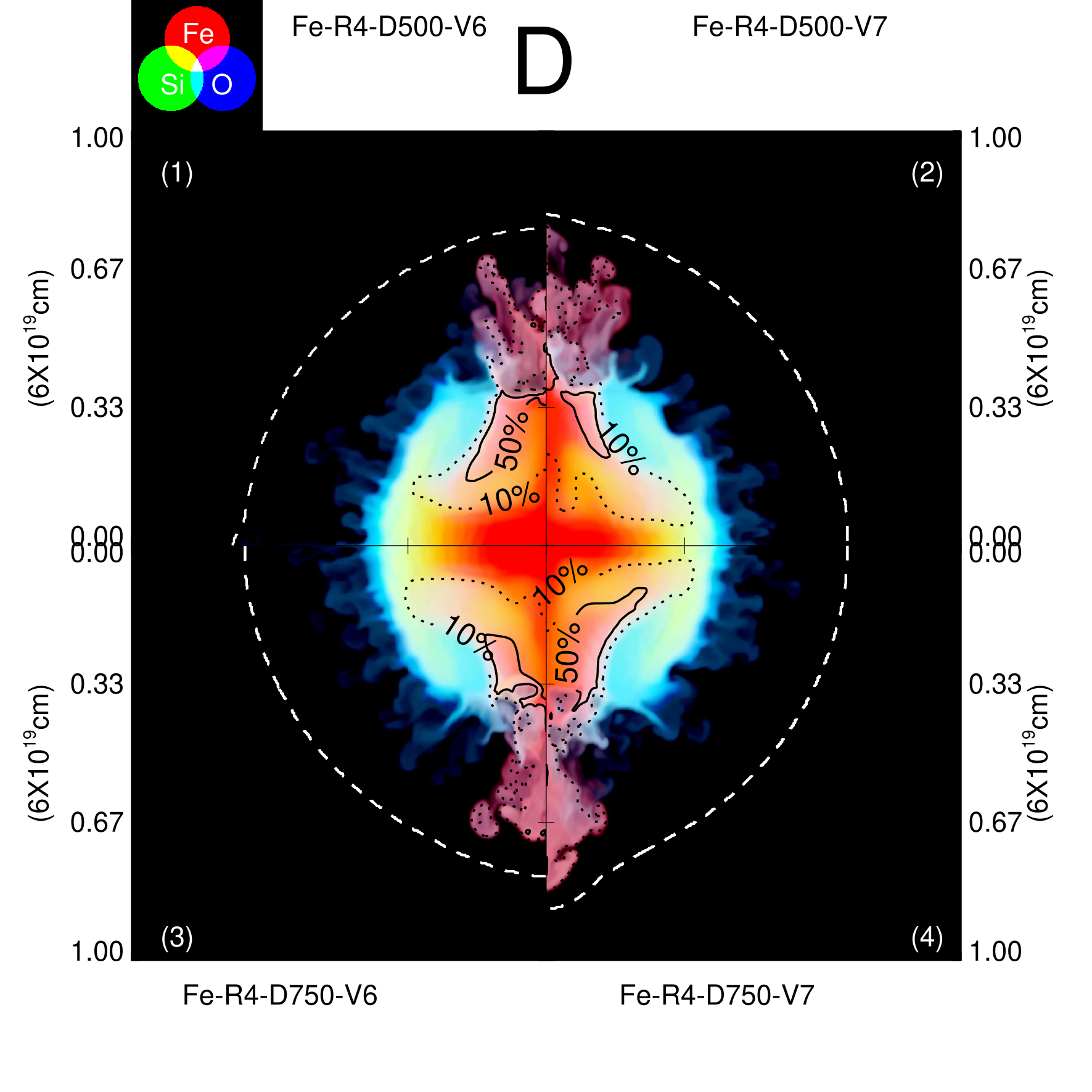}} 
   \caption{Color coded images of the logarithm of the ejecta mass fraction (> 10$^{-4}$) distributions at the end of the simulation ($t \approx 5000$ years) for different models (the model as presented in Table~\ref{tab:summary} is reported near each quadrant) in the ($x,0,z$) plane of $^{56}$Fe (red), $^{28}$Si (green) and $^{16}$O (blue). Black contours enclose the computational cells consisting of the original clump material by more than 50$\%$ (solid line) and 10$\%$ (dotted line). The white dashed line represents the projected position of the forward shock. See also online Movie 3, 4, 5 and 6 for an animated version.}
   \label{fig:RGB}
   \end{figure*}

Fig.~\ref{fig:RGB} (see also online Movie 3, 4, 5 and 6) shows the spatial distribution of $^{56}$Fe (red), $^{28}$Si (green) and $^{16}$O (blue) in the remnant's ejecta at $t=5000$ yr for different model setups (see Table~\ref{tab:summary}). Fig.~\ref{fig:RGB}A shows four models in which the large-scale clump is initially located at $D = 0.24$, i.e. between the $^{28}$Si and the $^{56}$Fe shells, with $r = 0.05$. In each of these cases, the clump can produce a remarkable jet-like structure of $^{28}$Si-rich ejecta. This structure protrudes beyond the SNR forward shock in all cases, except for model Si-R5-D500-V3 (Fig.~\ref{fig:RGB}A-1). We point out that the $^{28}$Si-rich collimated structure has overcome the $^{16}$O-rich layer of ejecta at this evolutionary stage. Moreover, we found out that after $\approx 350$ years, for models Si-R5-D500-V4 (Fig.~\ref{fig:RGB}A-2) and Si-R5-D600-V4 (Fig.~\ref{fig:RGB}A-4), this Si-rich jet-like structure has already overcome the $^{16}$O layer, a similar feature observed also in the northeast and southwest regions of Cassiopeia A (e.g. \citealt{DeLaney_2010, Milisavljevic_2013}).

Our simulations show that a velocity contrast $\chi_{\textrm{v}} > 2$ is necessary to allow the clump to overcome the high-density shell. The length of the jet-like feature depends on the initial density contrast of the clump, as it can be seen in quadrant (1) of Fig.~\ref{fig:RGB}A ($\chi_{\textrm{n}} = 500$, $\chi_{\textrm{v}} = 3$), where the position of the bow-shock of the clump has its lowest value ($4.12\times 10^{19}$ cm; $\approx 13$ pc) among the cases here considered. The shape of the jet-like structures appears somehow "curly", with the exception of Si-R5-D600-V3 (quadrant 3 of Fig.~\ref{fig:RGB}A) that has a more regular shape and is more collimated along the z-axis. The clump has been strongly eroded at this time.

The final distribution of $^{56}$Fe, $^{28}$Si and $^{16}$O in the remnant in the case of $D = 0.15$ (clump entirely made of $^{56}$Fe) and $r = 0.05$ is shown in Figs.~\ref{fig:RGB}B, C. In both cases the clump produces a $^{56}$Fe-rich region along the z-axis, extending from the center of the remnant to the intershock region where it mixes with the $^{16}$O layer. We found that the clump can protrude the forward shock only if its velocity contrast is $\chi_{\textrm{v}} \geq 6$, that is a factor of 2 higher than that obtained for $D = 0.24$. This is because, for $D=0.15$, the clump is located in an inner and slower region (due to the ejecta's linearly increasing velocity profile described in Section~\ref{method}), therefore a higher velocity contrast is needed with respect to the case with $D=0.24$. Moreover, for $D = 0.24$, the clump is at the inner border of the high-density shell, so the interaction with the shell occurs at the early stages of evolution, when the clump to shell mean density ratio (in the following $\eta$) is $\eta \geq 14$. On the other hand, for $D=0.15$, the clump evolves in a lower-pressure region, therefore the clump undergoes a significant expansion before interacting with the dense shell and $\eta \leq 0.25$ at the interaction. This effect causes the clump to struggle to overcome the dense shell, leading to a wider jet-like structure with respect to cases with $D=0.24$. We also found that the initial clump size ($r$) does not affect the mixing of $^{56}$Fe in the outer layers (see Fig.~\ref{fig:RGB}D) within the range of values explored. 

In the cases explored, we observed that the clump pushes the heavy elements (especially $^{56}$Fe and $^{28}$Si) into the outer and high-speed regions of the ejecta, as it has been observed in other core-collapse SNe such as SN1987A~\citep{Haas_1990, Utrobin_1995}, Cassiopeia A~\citep{Grefenstette_2014,Grefenstette_2017, Siegert_2015}, Vela~\citep{Garcia_2017} and, very recently, N132D~\citep{Law_2020}. According to the exploration of the parameter space we have performed, the average chemical stratification of the ejecta is roughly preserved in regions not affected by the large-scale anisotropy (e.g. x-axis in Fig.~\ref{fig:RGB}). In these regions, the evolution of the ejecta is similar to that described in section~\ref{subsec:spher} and the forward shock reaches approximately the same distances in all the cases explored (the differences shown in quadrants (2) and (4) of Fig.~\ref{fig:RGB}B and C are due to slightly different simulation times, of the order $\approx$ 10 years).

   \begin{figure}
   \centering
   \includegraphics[width=\hsize]{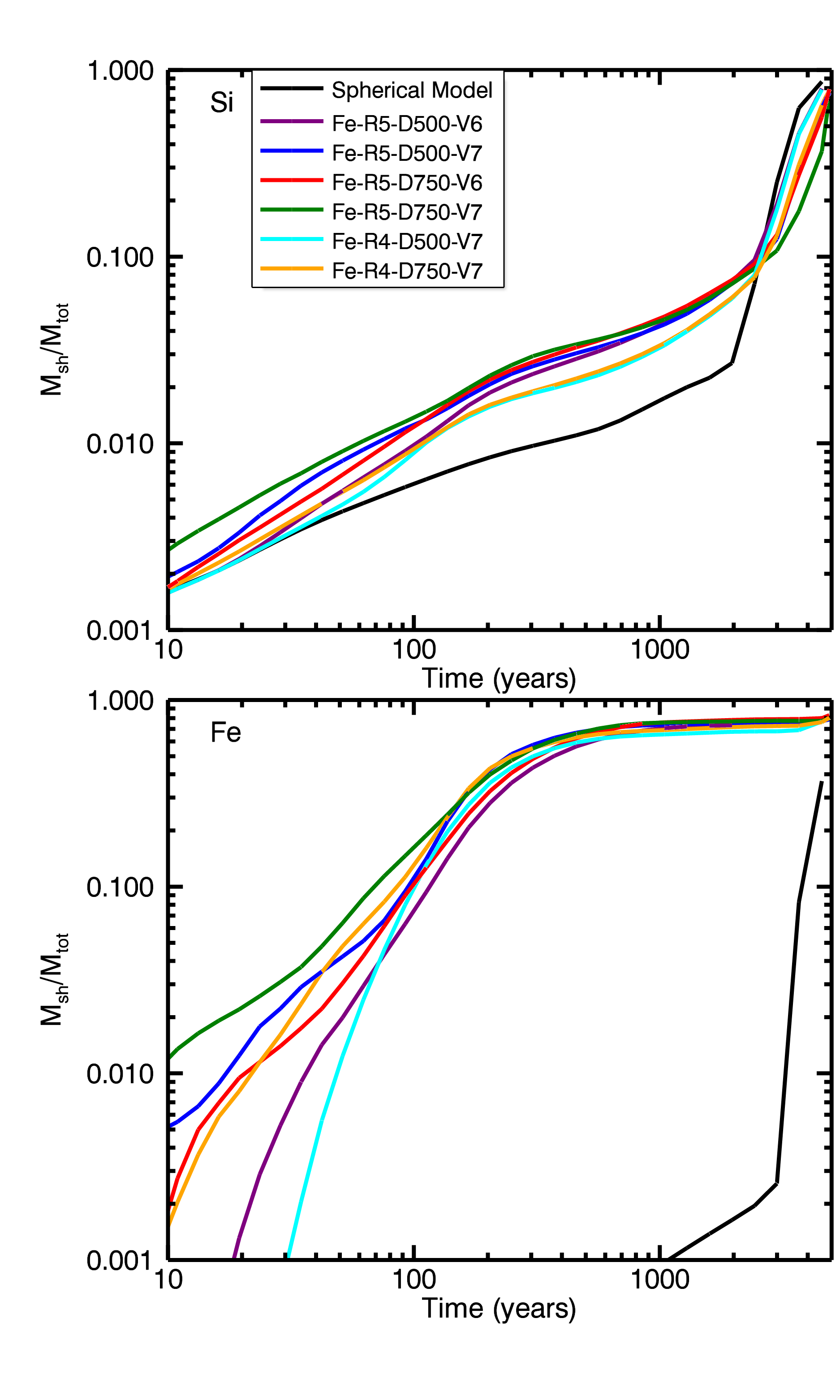}
      \caption{Mass of shocked $^{28}$Si (upper panel) and $^{56}$Fe (lower panel) vs. time for models assuming a clump initially located at $D = 0.15$ and characterized by different parameters (see Table~\ref{tab:summary}). The shocked mass is normalized to the total mass of the relative element.}
         \label{fig:plot_shock_Fe}
   \end{figure}
   
\begin{figure}
   \centering
   \includegraphics[width=\hsize]{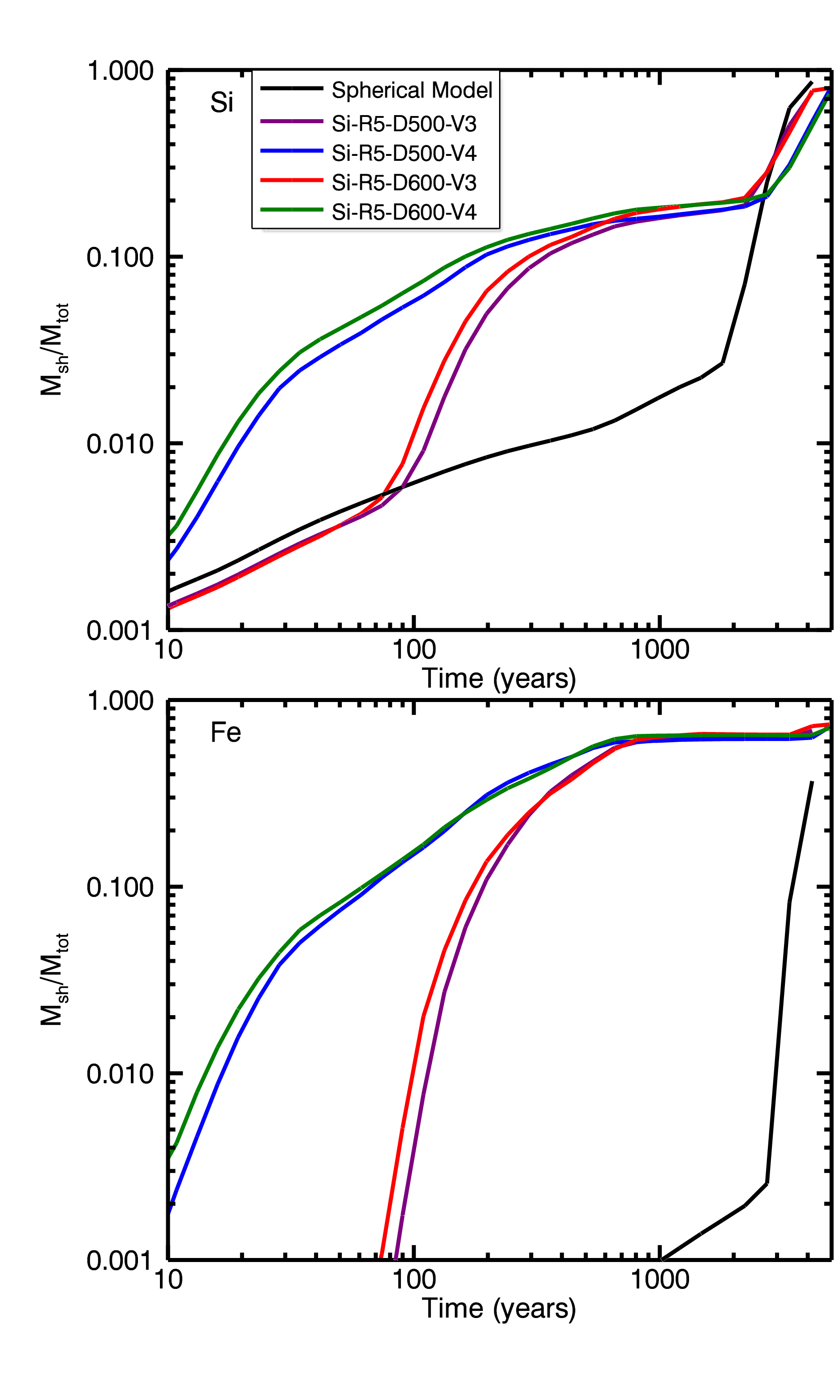}
      \caption{Same as Fig.~\ref{fig:plot_shock_Fe} but for models with a clump initially located at $D = 0.24$.}
         \label{fig:plot_shock_Si}
   \end{figure}

Shock-heated plasma emits X-ray thermal emission, especially through lines emission, giving us a valuable tool to measure the elemental composition and spatial distribution inside the remnant. From the simulations we derived the mass of shocked $^{56}$Fe and $^{28}$Si normalized to the value of the total mass of the respective element for the different models as a function of time (see Figs.~\ref{fig:plot_shock_Fe}-~\ref{fig:plot_shock_Si}). Both for $D=0.15$ and $D=0.24$, the final amount of shocked $^{56}$Fe is at least 2 times larger than that obtained for the case of a spherically symmetric explosion.
 
For $D = 0.15$ (see Fig.~\ref{fig:plot_shock_Fe}), during the first 300 years, the amount of shocked $^{56}$Fe depends both on the size and on the initial density contrast of the clump, increasing with size and contrast. However, after 3000 yr, the reverse shock interacts with the $^{56}$Fe shell even without any initial clump, therefore the amount of shocked $^{56}$Fe observed at the end of the simulations is not entirely due to the initial clump. As for the shocked $^{28}$Si, we found that the final amount of shocked mass does not show any strong dependence on the initial clump parameters and on its initial dimension and position. The main cause of this is the interaction of the reverse shock with the Si-rich shell that shocks a relevant part of the $^{28}$Si-rich material, regardless of the presence of the initial clump, thus making the contribution of the clump negligible.
For $D = 0.24$ (see Fig.~\ref{fig:plot_shock_Si}) instead, the amount of shocked $^{56}$Fe after 5000 yr is not strongly affected by the initial clump parameters. This is expected since the jet-like structure is mostly made of $^{28}$Si for models with $D = 0.24$, as shown in Fig.~\ref{fig:RGB}. On the other hand, the velocity contrast $\chi_{\textrm{v}}$ plays an important role in determining the beginning of the interaction between the reverse shock and the $^{56}$Fe-rich ejecta. We observed that, for $\chi_{\textrm{v}} = 4$, the reverse shock heats the $^{56}$Fe-rich material $\sim90$ yr earlier than the case with $\chi_{\textrm{v}} = 3$ (see upper panel of Fig.~\ref{fig:plot_shock_Si}). As seen for the $^{56}$Fe, we found however that the initial velocity contrast $\chi_{\textrm{v}}$ affects the timing of the interaction between the reverse shock and the $^{28}$Si-rich ejecta, only for $D = 0.24$.

\begin{figure}
   \centering
   \includegraphics[width=\hsize]{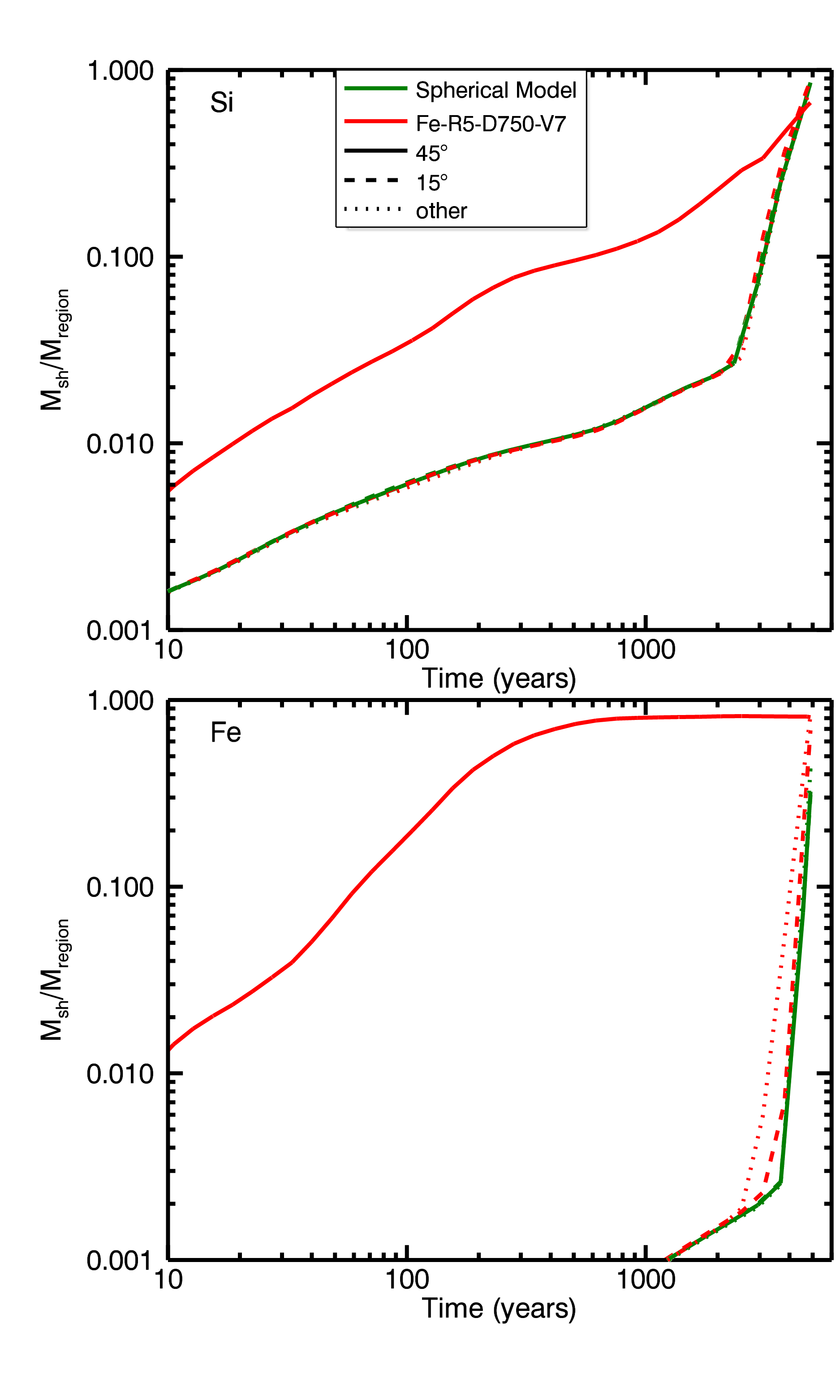}
            \caption{Mass of shocked $^{28}$Si (upper panel) and $^{56}$Fe (lower panel) vs. time for the spherically symmetric explosion (green lines) and for model Fe-R5-D750-V7 (red lines, see Table~\ref{tab:summary}). Solid and dashed lines show the shocked mass from cells within 45$^{\circ}$ of the z-axis and from 15$^{\circ}$ of the equatorial plane respectively, while the dotted lines show the mass from the others cells. The shocked mass is normalized to the total mass of the relative element in the region considered.}
         \label{fig:rot}
   \end{figure}

To deepen our understanding of the role played by the clump, we also computed the amount of shocked $^{28}$Si (upper panel of Fig.~\ref{fig:rot}) and $^{56}$Fe (lower panel of Fig.~\ref{fig:rot}) as a function of time in 3 different regions of the ejecta, namely: i) cells within $45^{\circ}$ of the z-axis (i.e., those enclosing the ejecta clump and mostly affected by its evolution) ii) cells within $15^{\circ}$ from the equatorial plane (i.e., those least affected by the clump) , and iii) all the other cells. The shocked mass is normalized to the total mass of the respective element in the corresponding region. We compared the spherically symmetric explosion (green lines) and model Fe-R5-D750-V7 (red lines, see Table~\ref{tab:summary}). For model Fe-R5-D750-V7, in the region within $45^{\circ}$ of the z-axis the mass of the $^{28}$Si (upper panel) has been shocked by a larger fraction, compared to the other regions, but only up to $t=4000$ yr. At odds with the $^{56}$Fe (lower panel), after $t=4000$ yr the reverse shock reaches the $^{28}$Si layer in the regions away from the jet-like structure making these regions the main contributors to the shocked mass of $^{28}$Si. Looking at the evolution of the shocked mass of $^{56}$Fe (lower panel) for model Fe-R5-D750-V7, the effect of the clump in cells within $45^{\circ}$ of the z-axis) stands out, while the other regions follow the same evolution as the spherically symmetric explosion. However, the presence of the clump influences the regions within $45^{\circ}$ of the equatorial plane shocking $40\%$ more mass of the region compared to the spherically symmetric explosion.

\subsection{Synthesis of X-ray emission}
 
In this paper we showed that the presence of an anisotropy can significantly influence the shape of the remnant leading to the formation of large-scale jet-like structures and protrusions of the remnant outline. The matter-mixing strongly depends on the physical and chemical characteristics of the initial anisotropy and it is now interesting to have a probe of such mixing by looking at the X-ray spectra. When the shock wave interacts with the matter, the latter is heated up to temperatures of several millions degrees, leading to X-ray emission. Thus, X-ray spectral analysis represents a powerful tool to appreciate different spectral signatures of the regions affected by an ejecta bullet and the rest of the evolving remnant between the forward and the reverse shock.

\begin{figure}
     \centering
     {\includegraphics[width=.5\textwidth]{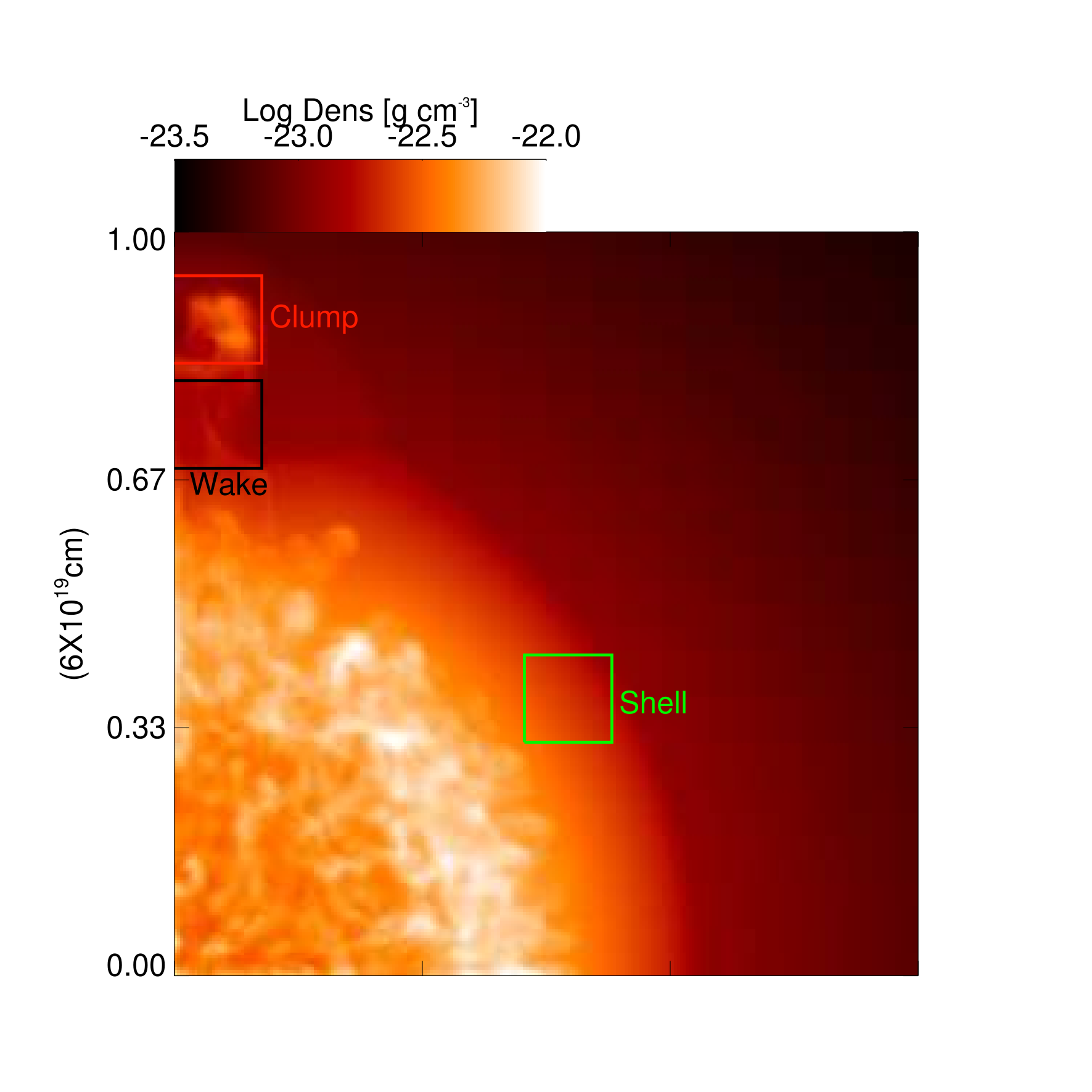}} \quad
     {\includegraphics[scale=0.35]{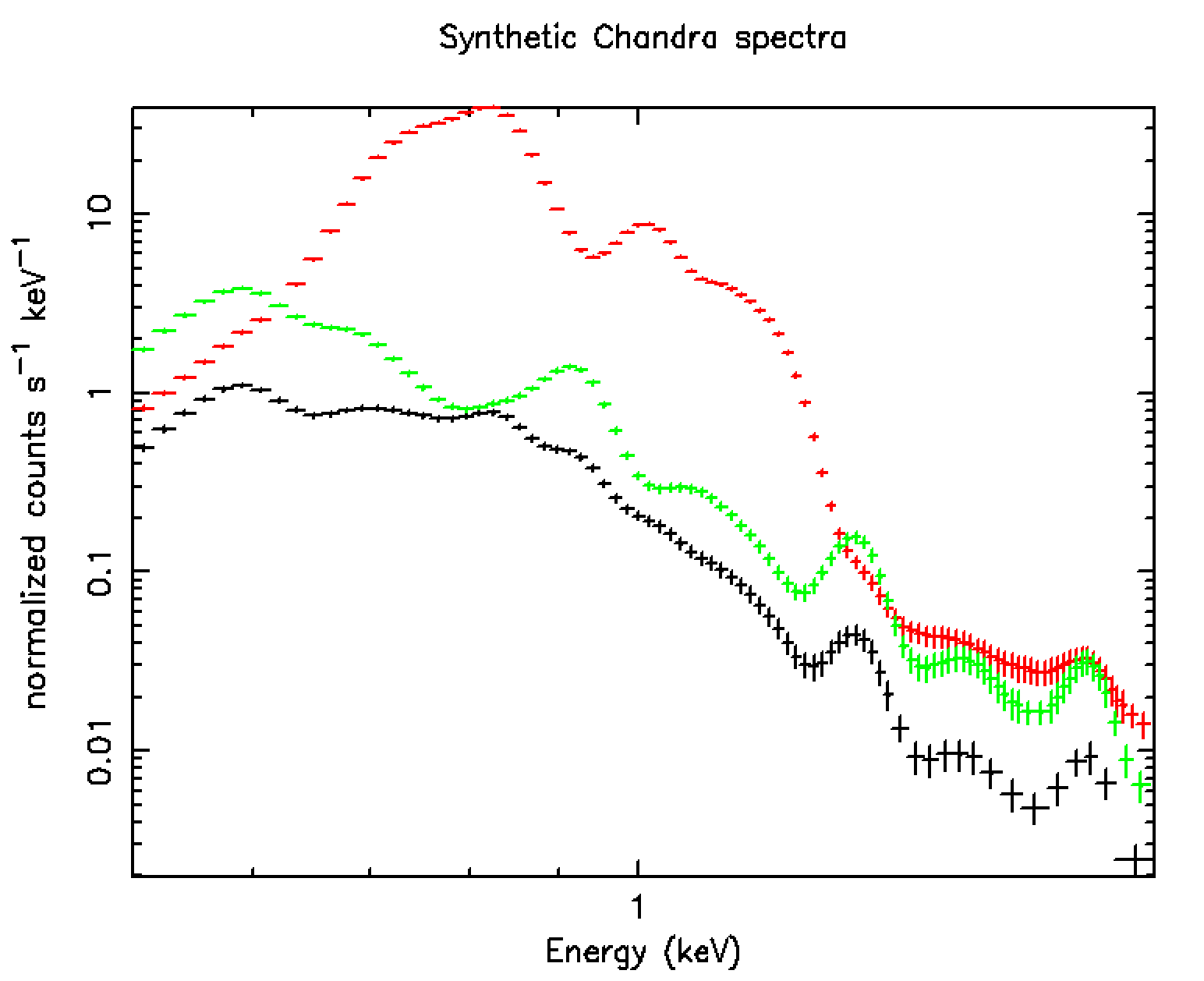}} \\
     \caption{Upper panel: density distribution map integrated along the line-of-sight (in this case the $y$-axis) of the run Fe-R5-D750-V7, showing the regions chosen for the spectral synthesis: red for the anisotropy protruding the remnant outline (the clump), black for the wake of the anisotropy and green for a region of the shell not affected by the anisotropy (the shell). Lower panel: X-ray \emph{Chandra} ACIS-S synthetic spectra in the 0.5-2 keV energy band for the clump (red), the wake (black), and the shell (green).}
     \label{fig:synthesis}
 \end{figure}

By adopting the approach described in \citealt{Greco_2020} (see also \citealt{Miceli_2019}), we self-consistently synthesized X-ray spectra from our simulations in a format virtually identical to that of \emph{Chandra} ACIS-S observations. The synthesis includes the deviations from equilibrium of ionization and from temperature-equilibration between electrons and protons, and takes into account the chemical composition of ejecta and ISM resulting from the simulations in each computational cell (see e.g. \citealt{Miceli_2019, Orlando_2019} for more details). For the spectral synthesis, we assumed a distance of 1 kpc and a column density $N_H=2\cdot$ 10$^{21}$ cm$^{-2}$ with an exposure time of 100 ks. As an example, we analyzed run Fe-R5-D750-V7 (see Table~\ref{tab:summary}) where a clear jet-like feature is evident at an age of $t=5000$ yr. We selected three regions identified by the red, black and green boxes in the upper panel of Fig.~\ref{fig:synthesis} corresponding to the ejecta bullet protruding the remnant outline (in the following the clump), the wake (namely the low-density strip-region behind the clump, already mentioned in Sect.~\ref{sec:results_anisotropy}), and an area of the shell not affected by the anisotropy, respectively. For a distance from the observer of 1 kpc, each of these regions corresponds to a box with angular side $l=2'$. Synthetic spectra extracted from the three regions are shown in the lower panel of Fig.~\ref{fig:synthesis}. 
 
 The main spectral difference between the spectrum extracted from the region enclosing the clump and the other spectra is the bright "false-continuum" emission due to the blending of Fe XIV-XXIV emission lines at energies around 1 keV. This result shows that the clump spectrum is dominated by the  emission of Fe-rich ejecta. Emission lines of other elements such as Mg ($\sim$ 1.3 keV) and Si ($\sim$ 1.8 keV) are fainter in the clump spectrum, since the protrusion of the Fe-rich clump drags away lighter elements. Enhanced Ne (at $\sim0.92$ keV), Mg, and Si emission lines are visible in the spectrum of the shell region (green spectrum in Fig. \ref{fig:synthesis}). The wake spectrum (black spectrum in Fig. \ref{fig:synthesis}) shows bright Mg and Si emission lines, very low Fe emission, even lower than that in the shell region, and an overall lower emission measure. The emission is fainter because the density is low, as we can see in the integrated density distribution map displayed in the upper panel of Fig.~\ref{fig:synthesis}). Our analysis shows that the effects of matter mixing in the ejecta produced by the evolution of a post-explosion anisotropy lead to spectral features in the X-ray band which are detectable through a spatially resolved analysis of X-ray spectra of SNRs.
   
 \section{Summary and conclusions}\label{Conclusions}

We investigated the evolution of the ejecta in core-collapse SNRs from the onset of the SN to the full-fledged remnant. The aim was to analyze how matter mixing occurs during the remnant expansion and interaction with the CSM, how the various chemically homogeneous layers at the time of the explosion map into the resulting abundance pattern observed when the remnant is fully developed, and how the presence of large-scale anisotropies affects the evolution of these layers. To this end, we developed a 3D MHD model describing the evolution of a SNR starting soon after the SN explosion and following the interaction of the SN ejecta with the CSM for 5000 years. As initial conditions for the SNR simulations, we adopted the output of a spherical SN explosion model by~\citealt{Ono_2020} (see also \citealt{Ono_2013}), soon after the breakout of the shock wave at the stellar surface (about 17 hours after the core-collapse). More specifically we considered the case of a $19.8$ M$_{\odot}$ progenitor RSG which exploded as SN with an injected energy of $2.5 \times 10^{51}$ erg. The initial ejecta structure was modeled by including small-scale clumping of material and a larger-scale anisotropy. The simulations are multi-species to trace the life cycle of elements during the whole evolution, from the nuclear reaction network of the SN, to the enrichment of the CSM through mixing of chemically homogeneous layers of ejecta.

To ascertain the effect on the matter-mixing of the spherically symmetric expansion of the remnant, we first explored a case with  small-scale clumpy structures of ejecta without considering any large-scale anisotropy. The ejecta soon after the SN explosion show the original onion-like structure, with  elements arranged in concentric shells. Although, after 5000 years of evolution, the onion-like structure is maintained due to the homologous expansion of the unshocked layers of the ejecta, we observed a significant matter mixing in the inter-shock region of the remnant, as also expected from previous studies (e.g. \citealt{Gull_1973, Chevalier_1976, Fryxell_1991, Chevalier_1992}; see  Fig.~\ref{fig:spher_massfrac}). This mixing is caused by the development of HD instabilities generated by the interaction of the ejecta with the CSM. However, this mixing is not able to cause a spatial inversion between chemical layers, leaving the initial chemical configuration of the ejecta roughly preserved.

We then performed a set of simulations to study the evolution of a post-explosion large-scale anisotropy (the clump) in the ejecta and its effects on the matter-mixing, investigating the role of its initial parameters (position, dimension, density and velocity contrasts). We explored primarily two cases: 1) a clump initially located in the $^{56}$Fe/$^{28}$Si region ($D = 0.24$)  and 2) a clump in the $^{56}$Fe region ($D = 0.15$). In both cases, the clump is located behind a thick and dense shell of lower-$Z$ ejecta. This thick and dense shell requires a higher density contrast for the clump, compared to previous works~\citep{Miceli_2013, Tsebrenko_2015}, to reach the forward shock. The dense-shell structure could strongly depend on the progenitor model. This because different progenitors could produce different density profiles after the explosion and, therefore, we should obtain different initial conditions and the clump would evolve in a different environment. The progenitor discussed in this paper (a $19.8$ M$_{\odot}$ progenitor RSG) is a generic case of a RSG among the different progenitors simulated in~\cite{Ono_2020}. The magnetic field is not relevant for the overall evolution of the remnant (being the plasma-$\beta$ much larger than 1) but plays some role in reducing the fragmentation of the clump by the HD instabilities that would develop at its border~\citep{Fragile_2005, Shin_2008, Orlando_2008, Orlando_2012}.

From the first $\approx$10 years of evolution, the clump rapidly assumes an elongated structure and its outermost region becomes denser. It takes $\sim$100 yr for the clump to overcome the high-density shell of lower-$Z$ ejecta and start to interact with the reverse shock and the intershock region. After $\sim$1000 years the HD instabilities begin to erode the clump, by gradually fragmenting it, supported by the effects of compression and heating caused by the interaction of the clump with the reverse shock. We found that the initial position of the clump strongly affects the morphology of the clump at $t = 5000$ yr. A clump initially located at $D = 0.24$, i.e. attached to the high-density shell (see right quadrant of Fig.~\ref{fig:init_confr}), overcomes entirely the high-density shell and, at the end of the simulation, only a small fraction of the clump is left behind the reverse shock (e.g., top left panel of Fig.~\ref{fig:RGB} where contours enclosing cells with more than 10\% of the original clump material are shown). Nevertheless, a clump initially located at $D = 0.15$ expands significantly before interacting with the shell. This involves a minor density contrast of the clump at the interaction with the shell, and consequently a major difficulty to overcome the shell. This translates in a higher density contrast of the clump required to overcome the high-density shell of lower-$Z$ ejecta and a bigger fraction of the clump left behind the reverse shock at the end of simulation time (see top right and bottom left panels of Fig.~\ref{fig:RGB}). 

Furthermore, we found that for $D = 0.24$ a clump is able to protrude the forward shock with an initial velocity contrast $\chi_{\textrm{v}} = 3$, while for $D = 0.15$ $\chi_{\textrm{v}}>6$ is required. For a smaller clump ($r = 0.04$) only in model Fe-R4-D750-V7 ($\chi_{\textrm{n}}$ = 750, $\chi_{\textrm{v}}$ = 7) the clump is able to protrude the forward shock.

The chemical stratification in the ejecta is strongly affected by the clump, especially in the region along the z-axis. The initial onion-like structure is not preserved, as the clump, propagating through the remnant, forms a stream of $^{28}$Si($^{56}$Fe) ejecta, for $D = 0.24$ ($D = 0.15$), from the inner regions up to the intershock region. The clump makes its way out to the external layer of the remnant by piercing the chemically homogeneous shells and pushing the layers on the side of the stream. These streams cause a spatial inversion of the chemical layers, bringing the $^{28}$Si/$^{56}$Fe externally to the $^{16}$O shell (see Figs.~\ref{fig:RGB} and~\ref{fig:synthesis}). Where the effects of the clump's passage are not relevant, i.e. in the $(x,y,0)$ plane, the evolution of the chemical layer is analogous to that described by the spherically symmetric model (see Fig.~\ref{fig:rot}). The main effect of this inversion is on the amount of shocked mass of both $^{56}$Fe and $^{28}$Si as shown in Figs.~\ref{fig:plot_shock_Si},~\ref{fig:plot_shock_Fe} and \ref{fig:rot}. We found that a clump initially located in the $^{56}$Fe/$^{28}$Si region does not increase the final amount of shocked $^{28}$Si,  whereas it doubles the amount of shocked $^{56}$Fe. Although the density contrast ($\chi_{\textrm{n}}$) does not produce large effects, the velocity contrast ($\chi_{\textrm{v}}$) has an important role in determining the age of the shocking time of the elements: a higher contrast causes an earlier interaction. As a result, this will produce an earlier appearance of X-ray emission from shocked ejecta. In particular, we found that for a spherically symmetric expansion the $^{28}$Si ($^{56}$Fe) begin to be shocked after $\sim 2000$ yr ($\sim 3000$ yr), while the presence of the clump causes an early interaction of both elements with the reverse shock at most $\sim 100$ yr in advance. For a clump initially located in the Fe shell (i.e. $D = 0.15$), we found that the trend over time of the shocked $^{28}$Si mass is not affected by the presence of the initial large-scale anisotropy, while on the contrary the final amount of shocked $^{56}$Fe increase with increasing size, velocity and density contrast of the clump.

Our simulations allowed us to study the effect of a large scale anisotropy on the evolution and matter-mixing of the remnant's ejecta and provided a valuable tool to reproduce observable properties in SNRs. By synthesizing X-ray spectra from the model we found that it is possible to obtain a deeper level of diagnostics of ejecta inhomogeneities in CC-SNe by carefully comparing X-ray observations of SNRs with numerical simulations. Thus, our findings can be a useful guide in the interpretation of observations as, for instance, the Si-rich shrapnel protruding beyond the front of its primary blast shock wave in the Vela SNR~\citep{Garcia_2017} or the fast runaway knot with a significantly high Si abundance recently detected in N132D~\citep{Law_2020}.

\begin{acknowledgements}
We acknowledge that the results of this research have been achieved using the PRACE Research Infrastructure resource Marconi based in Italy at CINECA (PRACE Award N.2016153460). The PLUTO code is developed at the Turin Astronomical Observatory (Italy) in collaboration with the Department of General Physics of Turin University (Italy) and the SCAI Department of CINECA (Italy). The SN simulation has been performed with the FLASH code, developed by the DOE-supported ASC/Alliance Center for Astrophysical Thermonuclear Flashes at the University of Chicago (USA). The numerical computations of the SN simulation were carried out complementarily on XC40 (YITP, Kyoto University), Cray XC50 (Center for Computational Astrophysics, National Astronomical Observatory of Japan), HOKUSAI (RIKEN). SO, MM, FB and EG acknowledge financial contribution from the agreement ASI-INAF n.2017- 14-H.O, and partial financial support by the main-stream INAF grant “Particle acceleration in galactic sources in the CTA era”. This work is supported by JSPS Grants-in-Aid for Scientific Research “KAKENHI” Grant Numbers JP26800141 and JP19H00693. SN, MO and GF wish to acknowledge the support from the Program of Interdisciplinary Theoretical \& Mathematical Sciences (iTHEMS) at RIKEN (Japan). SN also acknowledges the support from Pioneering Program of RIKEN for Evolution of Matter in the Universe (r-EMU).
\end{acknowledgements}

      \begin{figure*}
   \centering
   {\includegraphics[width=.45\textwidth]{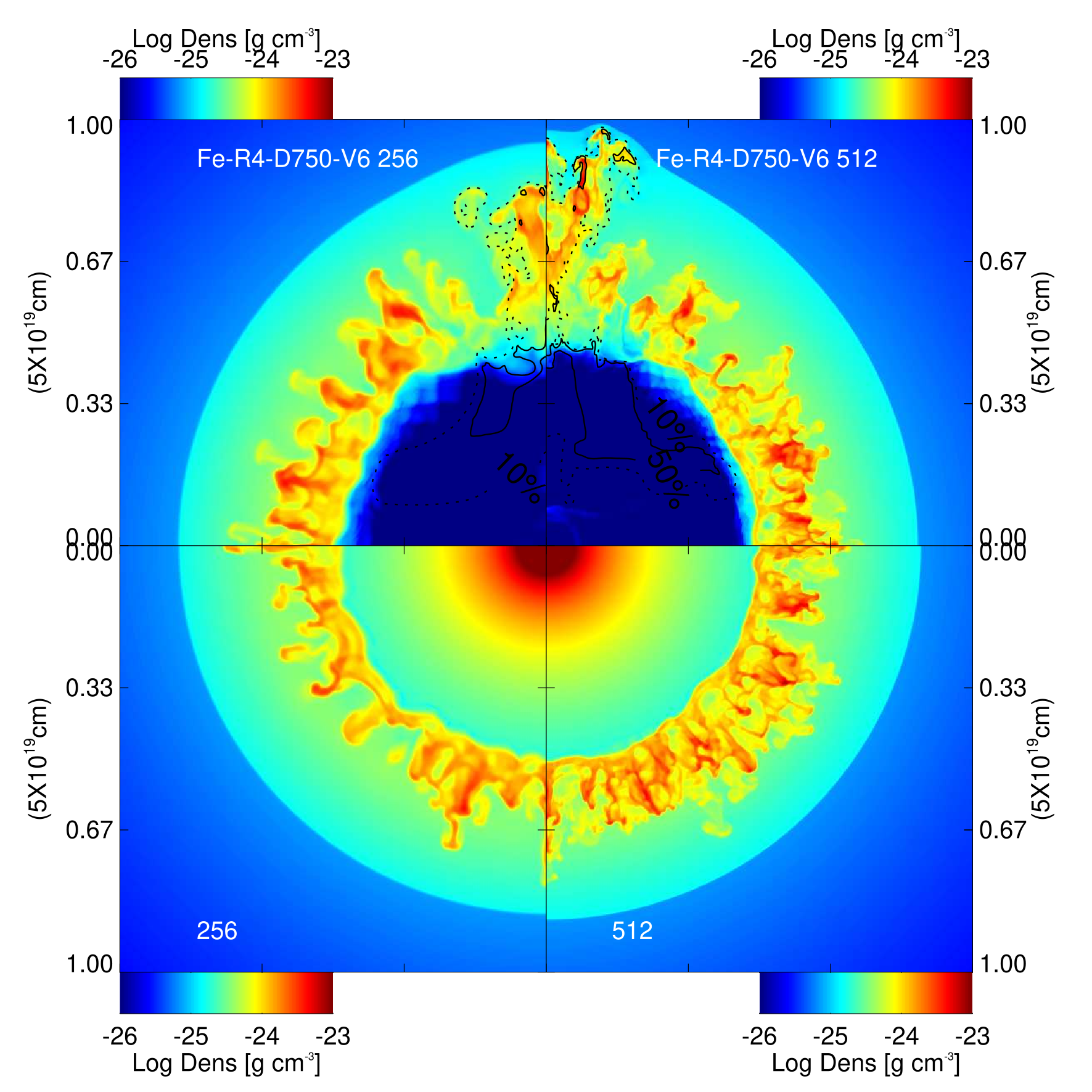}} \quad
   {\includegraphics[width=.45\textwidth]{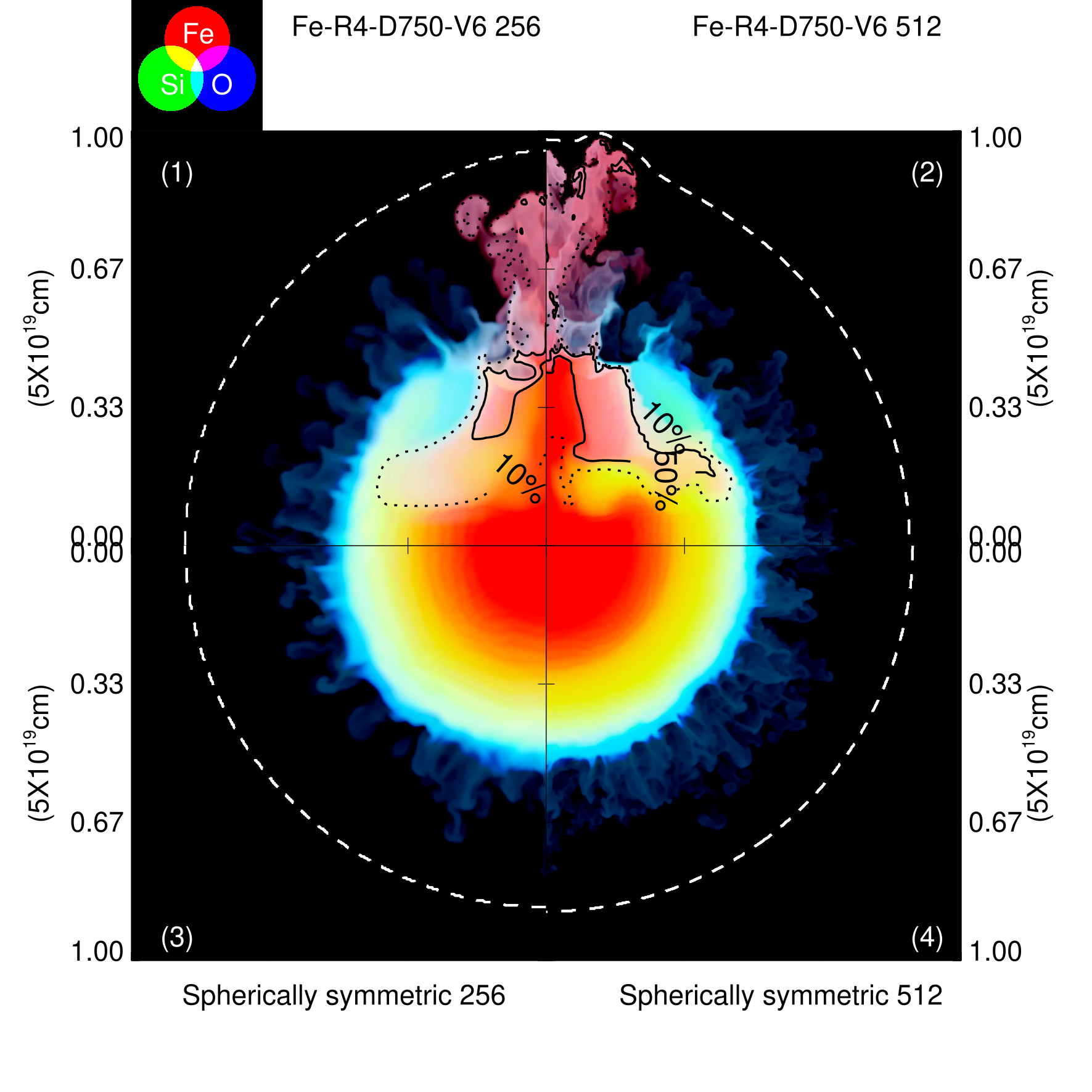}} \\
      \caption{Left panel: density distributions in the ($x,0,z$) plane at the end of simulation time for a SN explosion with a large-scale anisotropy (upper quadrants) (runs Fe-R4-D750-V6 and Fe-R4-D750-V6-512pt in Table~\ref{tab:summary}) and for a spherically symmetric explosion (lower quadrants), at low resolution ($256^3$ grid points; left quadrants) and high resolution ($512^3$ grid points; right quadrants). Right panel: same as left but for color coded images of the logarithm of the mass fraction distributions of Fe (red), Si (green), O (blue). In the upper quadrants of each panel, the black contours enclose the computational cells consisting of the original anisotropy material by more than 10$\%$ (dotted line) and 50$\%$ (solid line). The white dashed line represents the approximate position of the forward shock.}
         \label{fig:resolution_confr}
   \end{figure*}
   
         \begin{figure*}
   \centering
            {\includegraphics[width=.45\textwidth]{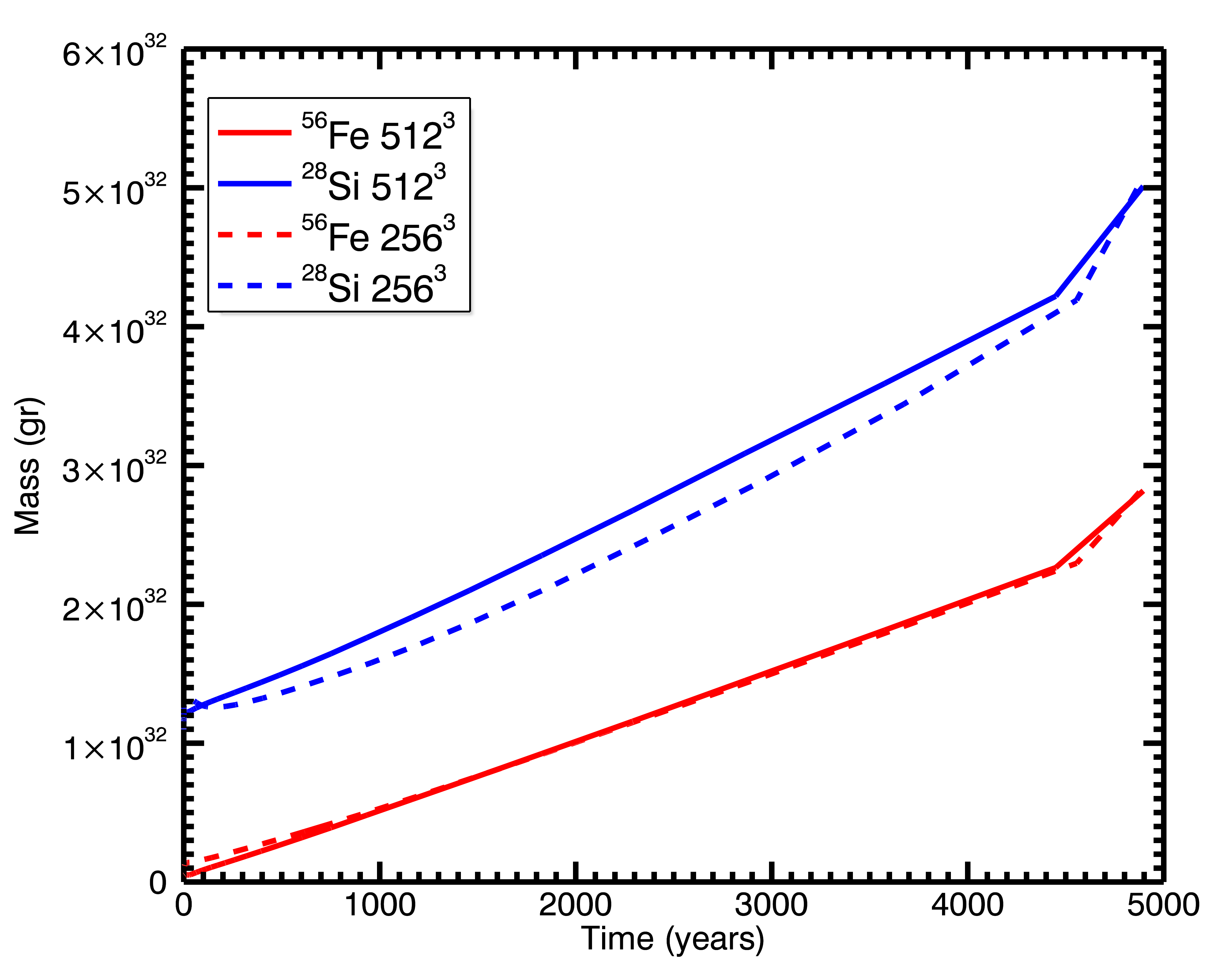}} \quad
            {\includegraphics[width=.45\textwidth]{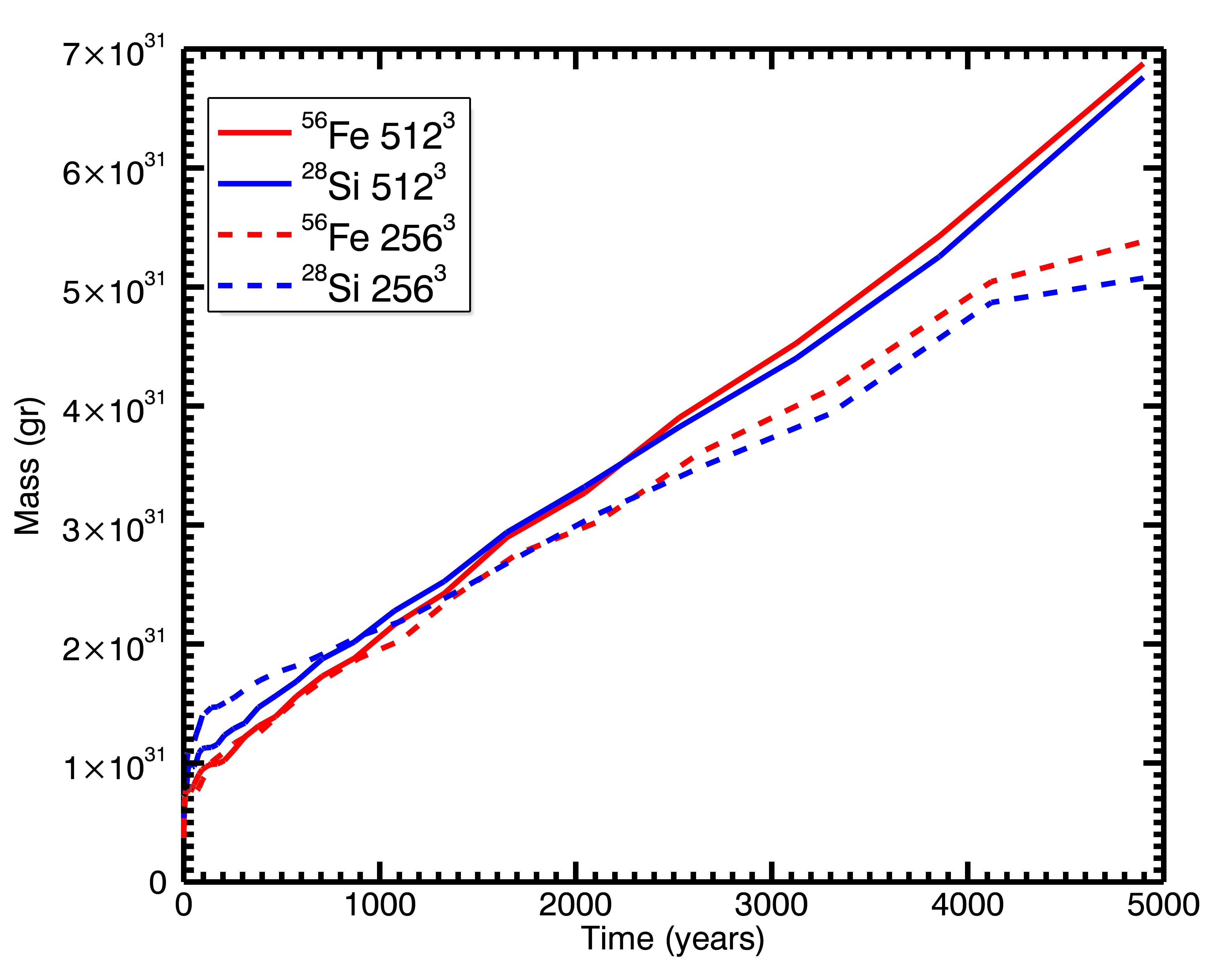}} \\
   \caption{Left panel: amount of $^{56}$Fe and $^{28}$Si mass vs. time in cells with a mass fraction between 10$\%$ and 50$\%$ for a spherically symmetric explosion at different resolution. Right panel: amount of $^{56}$Fe and $^{28}$Si mass vs. time in cells with a mass fraction between 10$\%$ and 50$\%$ and with anisotropy material between 10$\%$ and 90$\%$ for models Fe-R4-D750-V6 ($256^3$ grid points) and Fe-R4-D750-V6-512pt ($512^3$ grid points) (see Table~\ref{tab:summary}).}
   \label{fig:plot_mix}
   \end{figure*}

   \section*{Appendix A: Effect of spatial resolution}
\label{AppendixA}

   To ensure that the resolution used to perform the simulations ($256^3$ grid zones) is sufficient to catch the basic properties of the system evolution, we repeated simulation Fe-R4-D750-V6 but using a uniform grid of $512^3$ grid zones (hereafter simulation Fe-R4-D750-V6-512pt) and we did the same for the spherically symmetric explosion simulation. Left panel of Fig.~\ref{fig:resolution_confr} shows density distributions in the ($x,0,z$) plane at the end of simulation time for both models at different resolutions. Right panel of Fig.~\ref{fig:resolution_confr} shows the same as left but for color coded images of the logarithm of the mass fraction distributions of $^{56}$Fe, $^{28}$Si, and $^{16}$O. As expected, the spatial resolution mainly affects the structure of the RT instability: the RT fingers appear more extended and branched in the case of higher spatial resolution (see right quadrants in the panels of Fig.~\ref{fig:resolution_confr}). As a consequence, the spatial resolution changes the clump structure during its evolution. We also computed the distance of the center of mass of Fe-rich ejecta from the origin of the explosion for models Fe-R4-D750-V6 and Fe-R4-D750-V6-512pt, obtaining a value of $2.9 \cdot 10^{19}$ cm and $3.2 \cdot 10^{19}$ cm respectively, thus showing a slight discrepancy, of the order of 10\%, in the distribution of $^{56}$Fe in the two cases. We checked that simulations Fe-R4-D750-V6 and Fe-R4-D750-V6-512pt yield similar results in terms of the amount of shocked material. We were mainly interested in the mixing between different chemical layers, we therefore computed the amount of $^{56}$Fe and $^{28}$Si within the cells where the mixing is more relevant for both spherical (left panel of Fig.~\ref{fig:plot_mix}) and non-spherical  (right panel of Fig.~\ref{fig:plot_mix}) models. The profiles shown in Fig.~\ref{fig:plot_mix} are in good agreement between the two models. This proves that, even if the small scale density structures developing during the propagation of the ejecta bullet depend on the spatial resolution, a resolution of $256^3$ uniform grid zones is able to capture the basic properties of the system evolution and of the matter-mixing among different chemical layers.

   \begin{figure}
   \centering
   \includegraphics[width=\hsize]{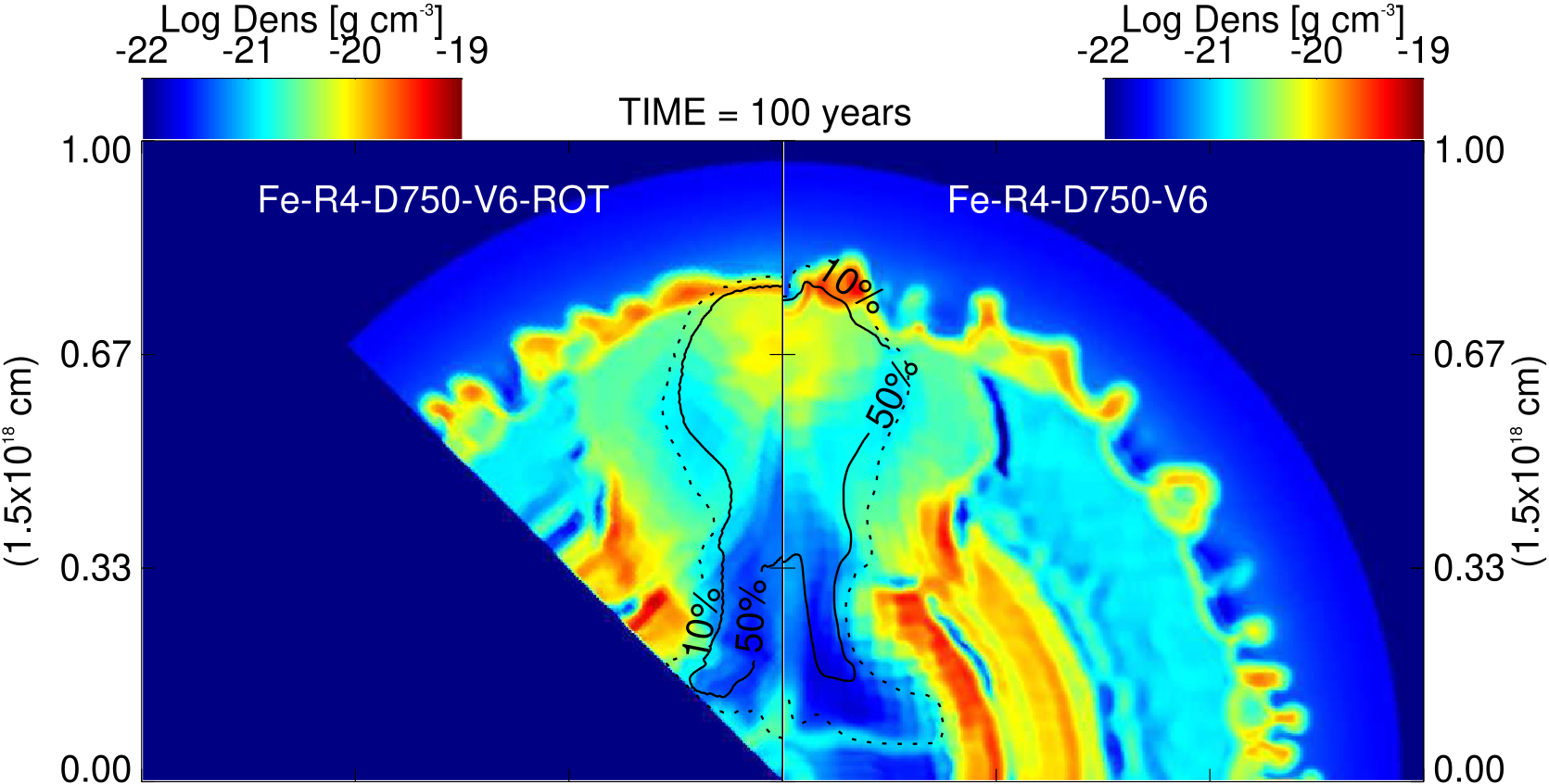}
   \includegraphics[width=\hsize]{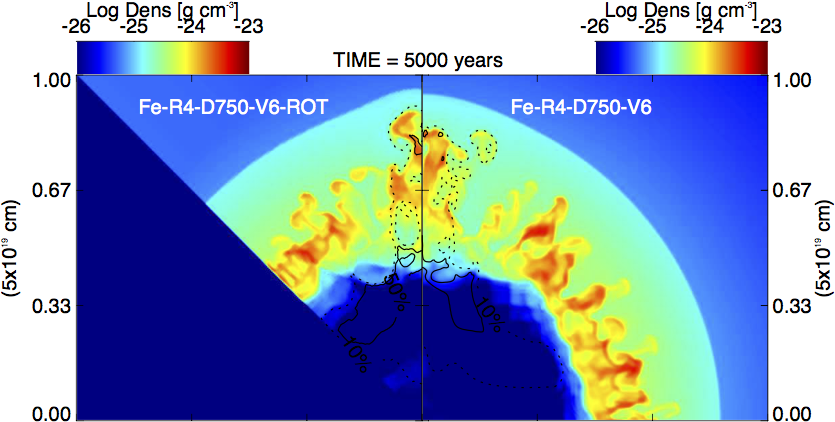}
      \caption{Density distribution maps in the (x,0,z) plane at $t \approx 100$ (upper panels) and $t \approx 5000$ years from the explosion (lower panels). On the right the model Fe-R4-D750-V6 (see Table~\ref{tab:summary}); on the left, same model with the initial position of the clump located at $45^{\circ}$ in the (x,0,z) plane, namely Fe-R4-D750-V6-ROT. On the left panels the (x,0,z) plane from the simulation is rotated to the z-axis for comparison, thus the lower blue triangle is not part of the computational domain. Note the different scales in the upper and lower panels. The contours enclose the computational cells consisting of the original anisotropy material by more than 50$\%$ (continuous line) and 10$\%$ (dotted line).}
         \label{fig:45_dens_RGB}
   \end{figure}
   
   \begin{figure}
   \centering
   \includegraphics[width=\hsize]{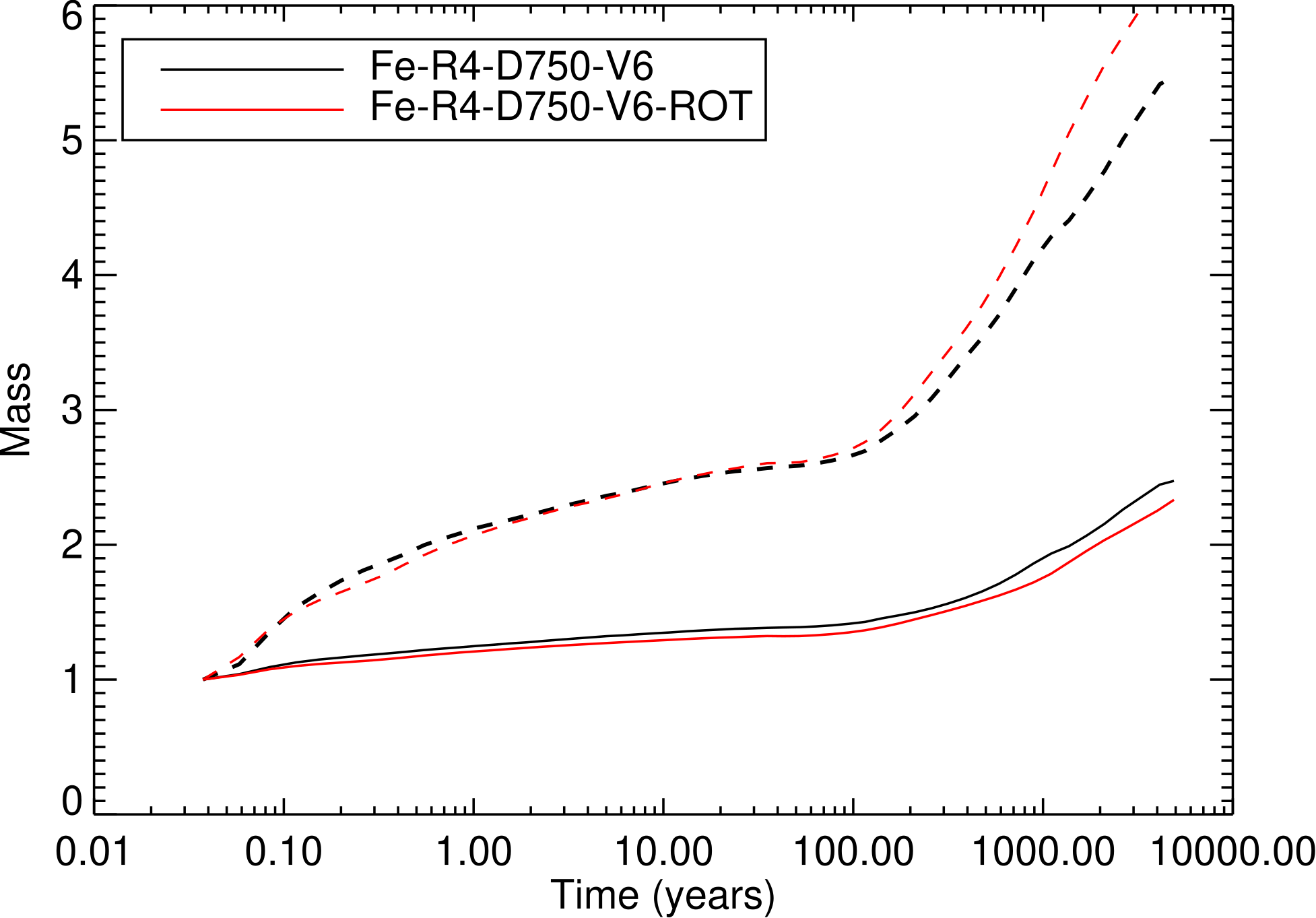}
      \caption{Mass vs. time (in logarithmic scale) for the two models compared in Fig.~\ref{fig:45_dens_RGB}. Total mass in cells with anisotropy material by more than 10$\%$ (continuous line) and in the range 10--90$\%$ (dotted line). In every case the mass is normalized to the value of the mass in the first step of the simulation.}
         \label{fig:45_mix_norm}
   \end{figure}
   
   \section*{Appendix B: Boundary effects}
\label{AppendixB}

In our numerical setup the large scale anisotropy, or clump, has been described as a sphere with its center located on the z-axis of the domain (see Sec.~\ref{sec:anisotropy}). In Section~\ref{sec:results_anisotropy} we have shown that the presence of this clump can cause a jet-like structure in the SNR that generates a strong mixing between different chemical layers. To ensure that the arise of this structure is not due to numerical artifacts caused by the proximity of the axis to the clump, we repeated simulation Fe-R4-D750-V6 placing the center of the clump at $\approx 45^{\circ}$ in the (x,0,z) plane (hereafter run Fe-R4-D750-V6-ROT). In Figure~\ref{fig:45_dens_RGB} we compare both models after $t \approx 100$ years of evolution and at the end of the simulation, namely $t \approx 5000$ years after the explosion. The evolution and final density distribution does not change significantly rotating the initial position of the clump out from the z-axis. More specifically, we noted that the evolution of the ejecta bullet is indistinguishable in the two cases during the first phase of the simulations (the first ~30 years), namely before the wake of the bullet reaches the (reflective) boundaries in run Fe-R4-D750-V6-ROT. In this latter case, the interaction of the wake with the boundaries produces reflection of material which propagates back into the domain. After $\sim 100$~yr (see upper panel in Fig.~\ref{fig:45_dens_RGB}), when the clump overcome the high-density shell of ejecta and start to interact with the inter-shock region, we observe a lower presence of HD instabilities in the model Fe-R4-D750-V6-ROT (left panel) and also some artifacts due to the reflection of material on the boundary during the expansion. For this reason, the mixing of anisotropy material in the two models is slightly different after $\sim 100$~yr of evolution (see Fig.~\ref{fig:45_mix_norm}). The perturbation of the inner portion of the remnant by the material reflected at the boundaries in run Fe-R4-D750-V6-ROT is the reason why we preferred to consider the bullet propagating along the $z$ axis rather than at $45^{\circ}$ angle from all axes. Nevertheless, even with the perturbations in run Fe-R4-D750-V6-ROT, the evolution of the ejecta bullet in both simulations is very similar with no significant effects on the final remnant structure (see Sec.~\ref{sec:results_anisotropy}). In the light of this, we can affirm that the jet-like structures observed in the simulations discussed in the paper do not arise from numerical artifacts.

\end{document}